# Derivation of Einstein–Cartan Theory from General Relativity

Richard James Petti, ORCID: 0000-0002-9066-7751 21 April 2021

146 Gray Street, Arlington, Massachusetts 02476 U.S.A. E-mail: rjpetti@gmail.com

**Abstract**. This work derives the elements of classical Einstein–Cartan theory (EC) from classical general relativity (GR) in two ways. (I): Derive discrete versions of torsion (translational holonomy) and the spin-torsion field equation of EC from one Kerr solution in GR. (II): Derive the field equations of EC as the continuum limit of a distribution of many Kerr masses in classical GR. The convergence computations employ "epsilon-delta" arguments, and are not as rigorous as convergence in Sobolev norm. Inequality constraints needed for convergence restrict the limits from continuing to an infinitesimal length scale. EC enables modeling exchange of intrinsic and orbital angular momentum, which GR cannot do. Derivation of EC from GR strengthens the case for EC and for new physics derived from EC.

## Contents

# 1. Introduction

## 1.1. Evolution of Einstein–Cartan theory

In 1922 E. Cartan proposed extending general relativity (GR) by dropping GR's ad-hoc assumption that affine torsion is zero [Cartan 1922; Cartan 1923]. In 1929 Einstein wrote to Cartan that he "didn't at all understand the explanations you gave me [about the role of torsion in geometry]; still less was it clear to me how they might be made useful for physical theory." [Debever 1979].

Sciama and Kibble derived Einstein–Cartan theory (EC) from a variational principle. They identified the modified torsion with intrinsic angular momentum (a.m.). They argue that inclusion of translational symmetries in the gauge group is a strong argument for EC [Sciama 1962; Kibble 1961].

Adamowicz showed that GR plus classical matter with spin yields the same linearized equations for the metric as EC [Adamowicz 1975].

The first modern review of EC was published in 1976 [Hehl 1976a]. A later review article treats EC as a viable theory (it satisfies all empirical tests of GR), but as an unproven speculation because it lacks empirical validation beyond empirical tests of GR [Trautman 2006]. A book of readings on gauge theories of gravity states that EC and teleparallel gravity are the only known alternatives to GR that satisfy all empirical tests satisfied by GR [Blagojević and Hehl 2013]. Hehl claims that the COW experiment can in principle provide an empirical test of EC [Colella et al, 1975; Gronwald and Hehl 1996].

EC can model the exchange of classical intrinsic and orbital a.m., which GR cannot due to symmetry of its momentum tensor. We define intrinsic a.m. as a.m. on too small a length scale to be represented as orbital a.m. in a given model. The best example of classical intrinsic a.m. is turbulence, which consists of transport of orbital a.m. to increasingly smaller scales [Monin and Yaglom 1971]. Current theories of turbulence do not track a.m. below the Kolmogorov limit. Modeling classical intrinsic a.m. with torsion is a current topic of research on turbulence [Peshkov 2019a; 2019b]. Turbulence is omnipresent in cosmological models, so a master classical theory of spacetime should be able to model exchange of intrinsic and orbital a.m.

## 1.2. Summary of results

Part One of this work updates the derivation of (a) translational holonomy, (b) a discrete version of torsion, and (c) a discrete version of the spin-torsion field equation of EC from one Kerr solution in GR [Petti 1986].

Part Two of this work derives the field equations of EC as the continuum limit of a distribution of Kerr masses in classical GR with constant densities of mass and a.m. The convergence computations employ "epsilon-delta" arguments, and are not as rigorous as convergence in Sobolev norm. Inequality constraints needed for convergence restrict the limits from continuing to an infinitesimal length scale.

In stellar and cosmological models, net electric charge is normally much smaller than the gravitational parameters. The final step in Part One and Part Two is to compare our computed discrete holonomy and continuum torsion with accepted gravitational models where charge is absent. Therefore we set electric charge $q = 0$ in Section 3 Theorem 3-c, Section 3.2 equation (19), Section 3.3, equation (29), and in Appendix D, lines 76 and 79.

The computations use concepts from differential geometry (translational holonomy, definition of torsion in terms of translational holonomy, surgery on manifolds) and physics (statistical theory of fluid turbulence, discrete affine defects in crystal lattices) that are less common in gravitation research.

The main value of the derivation of EC from GR is that it strengthens the case for EC and for new physics that is derived from EC.

The remainder of this work has these main sections:

- Section 2: Translational holonomy and the discrete version of torsion around one Kerr mass.

- Section 3: Torsion in the continuum limit of a discrete distribution of Kerr masses.
- Section 4: Conclusion and discussion

Appendices: (references in the article to Appendix D appear as “Apx D”)

A) Rotational curvature and affine torsion as limits of holonomy

B) Proof of Theorem 1

C) Proof of Theorem 2

D) Executable computer algebra derivation of results

E) Roles of spacetime and fiber tensor indices

F) Mathematics of Connection on fiber bundles.

G) Generation of rotational and translational holonomy on flat manifolds

## 2. Translational holonomy and the discrete version of torsion around one Kerr mass

We compute translational holonomy and a discrete version of torsion of some closed loops around a Kerr mass. The only significant translational holonomy arises from a spacelike loop with equatorial (or partially equatorial) orientation; the translation is timelike and proportional to the a.m. of the Kerr solution. The computation of translational holonomy uses only the distant region of an exterior Kerr solution.

Section 2.1 computes the translational holonomy for equatorial loops. Section 2.3 shows that translational holonomy for other loops is sufficiently small that the limit of torsion per unit area of the loop approaches zero.

Use the Kerr-Newman solution in Boyer-Lindquist coordinates [Misner et al 1973; O’Neill 1984].

(1) $$ds^2 = -\frac{\Delta}{\rho^2}[dt - a\sin^2(\theta)\,d\varphi]^2 + \frac{\sin^2(\theta)}{\rho^2}[(r^2 + a^2)\,d\varphi - a\,dt]^2 + \frac{\rho^2}{\Delta}dr^2 + \rho^2\,d\theta^2$$ (Apx D, d7)

where $0 \le \varphi \le 2\pi$, $0 \le \theta \le \pi$, and

(2) $$a := S/m$$

(3) $$\Delta := r^2 + a^2 - 2\,m\,r + q^2$$ (Apx D, e4)

(4) $$\rho^2 := r^2 + a^2\cos^2(\theta)$$ (Apx D, e5)

This metric is derived from a Minkowski metric using an orthogonal frame $FR_i{}^\mu$ and its inverse $FRI_\mu{}^i$. (Apx D, d6 and d2)

### *2.1. Translational holonomy of equatorial loops*

We shall calculate the translational holonomy of an equatorial loop of constant coordinate radius $r$. Define an equatorial loop $C(\sigma) : [0, 1] \to \Xi$ (spacetime) starting at $(0, r, \pi/2, 0) = p = C(0) \in \Xi$ with coordinates

(5) $$C(\sigma) = (0, r, \pi/2, 2\,\pi\,\sigma)$$ (Apx D, d48)

From this point forward, the computer algebra script enforces the inequality $r > m$ (Apx D, c48).

Develop $C(\sigma)$ into a flat Minkowski space $\Xi'$ with coordinates $(t', x', y', z')$. Start the development $C'(\sigma)$ in Minkowski space at point $p'$ with coordinates $(t'_0, x'_0, y'_0, z'_0)$.

The following reasoning determines values of parameters $c1$, $c2$, $k1$, and $k2$ that simplify the resulting expressions for holonomy and torsion.

a) In spacetime:
   - The radial coordinate for the equatorial loop must satisfy $r >> m$.
   - The curve parameter of the equatorial loop is $\sigma = \varphi/(2\,\pi)$.

b) In Minkowski space:
   - The shape, but not the location, of the developed curve affects the results because we use only the holonomy ("failure-to-close vector") of the developed curve.
   - The chosen starting point and initial direction of the developed curve transfer Kerr rotational symmetry to coordinates in Minkowski space. The curve forms a helix in 4-D, as shown in equations (10) and (13) (Apx D, Eq. d68).
   - The radius of the developed curve is a constant. We defined k2 to be this constant.
   - The timelike component of the tangent vector is a constant. the We defined k1 to be this constant.

c) Map of tangents from spacetime to Minkowski space:
   - The isometric tangent map (from tangents at the starting point of the loop in spacetime to tangents at the starting point of the developed curve in Minkowski space) is essentially determined. Any remaining arbitrariness of coordinates is used to transfer the Kerr rotational symmetry to the curve in Minkowski space.
   - We set c1 and c2 equal to the two most complicated repeated coordinate expressions in Tanmap to get the simple representation of Tanmap in Eq. (6) (Apx D, d59).

### *2.2. Development of equatorial spacetime loop into Minkowski space*

Define an isometry Tanmap: $T_p\Xi \rightarrow T_{p'}\Xi'$ from spacetime tangents at $p \in \Xi$ to the Minkowski tangents at $p' \in \Xi'$ defined by the matrix of partial derivatives

$$\text{(6)} \qquad \frac{\partial(t', x', y', z')}{\partial(t, r, \theta, \varphi)} = \begin{array}{c|cccc} & t & r & \theta & \varphi \\ \hline t' & \frac{c2}{\nu} & 0 & 0 & \frac{c1}{\nu} \\ x' & 0 & 1 & 0 & 0 \\ y' & -\frac{c1}{\nu} & 0 & 0 & -\frac{c2}{\nu} \\ z' & 0 & 0 & 1 & 0 \end{array} \qquad \text{(Apx D, d59)}$$

where parameters c1, c2, and $\nu$ enable writing the tangent map in a simple form.

$$\text{(7)} \qquad c1 := \frac{a}{r^2}\left(-r - m + \frac{q^2}{r}\right) \qquad c2 := -\frac{\Delta^{1/2}}{r} \qquad \nu := (c2^2 - c1^2)^{1/2} \qquad \text{(Apx D, c43)}$$

Since $|c2| > |c1|$, $\nu$ is real. Below is an exact closed-form expression for $\nu^2$.

$$\text{(8)} \qquad \nu^2 = 1 - \frac{2\,m}{r} + \frac{q^2}{r^2} - \frac{2\,a^2\,m}{r^3} + \frac{2\,a^2\,q^2 - a^2\,m^2}{r^4} + \frac{2\,a^2\,m\,q^2}{r^5} - \frac{a^2\,q^4}{r^6} \qquad \text{(Apx D, d43)}$$

If $|a|, |q|, m \ll r$, then

$$\text{(9)} \qquad \nu \approx 1 - m/r \approx 1$$

Theorem 1 presents the translational holonomy (the "failure-to-close vector") of the development of an equatorial loop into Minkowski space.

**Theorem 1:** The development of the equatorial loop C into the Minkowski manifold $\Xi'$ yields the curve

$$C'(\sigma) = (t'(\sigma), x'(\sigma), y'(\sigma), z'(\sigma))$$

which is given in Minkowski coordinates by (Apx D, d68):

$$\text{(10)} \qquad \begin{cases} t'(\sigma) = k1\,\sigma + t'_0 \\ x'(\sigma) = k2\,\cos(2\,\pi\,\nu\,\sigma) + x'_0 \end{cases} \qquad \text{(Apx D, d68)}$$

$$y'(\sigma) = k2 \sin(2 \pi \nu \sigma) + y'_0$$
$$z'(\sigma) = z'_0$$

Parameters $k1$ and $k2$ enable expressing the equatorial loop in a simple form.

(11) $$k1 = \frac{2 \pi a}{\nu r^4} (3 m r^3 - 2 q^2 r^2 + a^2 m r - a^2 q^2)$$ (Apx D, e46)

(12) $$k2 = (r^4 - a^2 m r + a^2 q^2) \frac{\Delta^{1/2}}{\nu^2 r^4}$$ (Apx D, e47)

$-4 \pi k2 \nu$ is the radial acceleration of the equatorial loop. (Apx D, d52)

$C(\sigma)$ is a closed circular arc in spacetime, but its development $C'(\sigma)$ into Minkowski space does not close unless $\nu = 1$. If $m, a, q << r$, then $\nu \approx 1$ and the slope $dt'/d\sigma = k1$ is proportional to parameter $a$ = a.m./mass.

The proof consists of finding an exact solution of equation (A - 1) in Apx A (shown in Minkowski coordinates in Apx D, line d62). The computation uses connection coefficients that have spacetime indices for the first index (the direction of differentiation), and orthonormal frame indices for the second and third indices. See Apx B for the details of the proof.

The translational holonomy ("failure-to-close vector") of the equatorial loop $C$ is $C'(1) - C'(0)$. Expressed with the orthonormal coordinates on Minkowski space Ξ', this is:

(13) $$\text{translational holonomy} = \begin{bmatrix} -k1 \\ k2 (\cos(2 \pi \nu) - 1) \\ k2 \sin(2 \pi \nu) \\ 0 \end{bmatrix}$$ (Apx D, d72)

### *2.3. Translational holonomy of spacelike polar loops*

Consider a circular spacelike loop that passes through the north and south poles of a rotating Kerr mass.

(14) $$C(\sigma) = \begin{cases} (0, r, 2 \pi \sigma, 0), & \text{for } 0 \le \sigma \le 0.5 \\ (0, r, 2 \pi (1-\sigma), \pi), & \text{for } 0.5 \le \sigma \le 1 \end{cases}$$

Develop this curve into the Minkowski space Ξ'.

**Theorem 2:** The translational holonomy of the polar loop $C$ is zero.

Proof: See Apx C.

Theorems 1 and 2 give us the translational holonomy for two closed spacelike curves with a sufficiently large constant radius around a Kerr mass. The holonomy around any circular spacelike planar loop of constant radius is, to linear approximation, $\cos(\theta)$ times the holonomy of the equatorial loop, where $\theta$ is the angle between the axis of the rotating object and the normal vector to the plane of the loop.

### *2.4. Torsion around a Kerr mass*

Torsion is the limit of translational holonomy around a loop divided by the area enclosed by the loop. We use as the area of an equatorial loop the area as seen by an observer at infinity; in these coordinates, the area is $\pi r^2$. This choice reflects what an observer would measure far from the Kerr mass, and it avoids computing area inside the loop. So the discrete version of torsion around an equatorial loop, using equation (13) and expressed in the orthonormal frame in Minkowski space, is:

$$\text{(15)}\qquad \frac{\text{translational holonomy}}{\text{area of loop}} = \begin{bmatrix} -\,k1 / (\pi\, r^2) \\ k2\,(\cos(2\,\pi\,\nu) - 1) / (\pi\, r^2) \\ k2 \sin(2\,\pi\,\nu) / (\pi\, r^2) \\ 0 \end{bmatrix} \qquad \text{(Apx D, d77)}$$

It is convenient to replace the frequency $\nu$ with the variable $\alpha = 1 - \nu \approx 0$ and to expand in terms of $m$, $r$, $a$, and $q$ to get (Apx D, d78):

$$\text{(16)}\qquad \frac{\text{translational holonomy}}{\text{area of loop}} = \begin{bmatrix} -\,\dfrac{2\,a\,(3\,m\,r^3 - 2\,q^2\,r^2 + a^2\,m\,r - a^2\,q^2)}{(1-\alpha)\,r^6} \\ \dfrac{(\cos(2\,\pi\,\alpha)-1)(r^4 - a^2\,m\,r + a^2\,q^2)(r^2 - 2\,m\,r + q^2 + a^2)^{1/2}}{\pi\,(1-\alpha)^2\,r^6} \\ \dfrac{\sin(2\,\pi\,\alpha)(r^4 - a^2\,m\,r + a^2\,q^2)(r^2 - 2\,m\,r + q^2 + a^2)^{1/2}}{\pi\,(1-\alpha)^2\,r^6} \\ 0 \end{bmatrix}$$

From one Kerr solution in GR, we have derived translational holonomy and the discrete version of torsion = (translational holonomy) / (area of loop). All results obtained up to this point are derivable from the Kerr solution assuming only that $r$ is large enough compared to $a$, $q$ and $m$.

## 3. Torsion in the continuum limit of a discrete distribution of Kerr masses

The objective of section 3 is to prove this theorem:

**Theorem 3:** Given a highly symmetric distribution of discrete Kerr masses in GR, with constant densities of mass and a.m., with no initial relative motion,

a) The distribution of Kerr masses converges to a continuum solution if the inequality constraints in Table 1 below are satisfied.

b) The continuum limit of this distribution has constant densities of mass and intrinsic a.m.

c) The solution has translational holonomy. With constraints, $|a| \ll r$, $m \ll r$, and $q = 0$, the discrete version of torsion converges to torsion that satisfies the spin-torsion relationship of EC (Apx D, d76 and d93).

### *3.1. Variables in the continuum model*

Specify a sequence of distributions of Kerr masses that converges to a continuum with constant mass density, spin/mass, and charge density.

- $r$ = average half-distance between centers of Kerr masses as seen by an observer at infinity. '$r$' is the radius of the equatorial loop used to compute holonomy.

  The cross-sectional area assigned to loop of radius $r$ around each Kerr mass is $\pi\, r^2$, and the volume is $4/3\pi\, r^3$. When we derive the discrete version of torsion ( = (translational holonomy)/area), the accuracy of the area factor is not significant because both the numerator and denominator are multiplied by approximately the same factor.

- $md$ = mass density in the continuum model. This density is approximately $md = m/(4/3\pi\, r^3)$.

  Computations use $md = 4/3\pi\ md$ to prevent proliferation of factors of $4/3\pi$. At the end of the derivation, we revert to the mass density $md$ in equation (29).

- qd = electric charge density in the continuum model. This density is approximately $qd = q/(4/3\pi\ r^3)$.

  Computations use $qd = 4/3\pi\ q_{density}$ to prevent proliferation of factors of $4/3\pi$. For comparing with standard gravitational results, we set $q = 0$.

### *3.2. Inequality constraints*

The distribution of Kerr masses in Theorem 3converges to EC when the inequality constraints in Table 1 are satisfied.

**Table 1: Inequality constraints on parameters of the model**

| Restriction | Rationale |
|---|---|
| (17) $m << r$ | This constraint ensures that the loops used to compute holonomy are far from the central region of the Kerr mass so m/r is small in the convergence computations. In Appendix D, this inequality first enters on line c48.<br>From $m << r$ and the definition $md = m/r^3$, the inequality $md\ r^2 << 1$ follows. |
| (18) $a^2 << r^2$ | This restricts the result to classical rotating objects. As an extreme case, a distribution of 100% spin-polarized neutrons would have $a \approx 5\ 10^{19}$ in Planck units. Such a model satisfies $\|a\| << r$ if $r > 10^{22}$, that is, $r > 10^{-12}$ m.<br>If $a = spin/mass$ has a fixed value, then as 'r' gets smaller, spin and mass of each Kerr mass decreases as $r^3$. This inequality prevents taking the limit $r \to 0$.<br>This inequality enters the computation in equation (9) (Apx D, c83). |
| (19) $\|qd\|\ r^2 << md$ | $4/3\ \pi\ qd$ is net charge density. qd enters holonomy in the form $qd^2\ r^2 + md$. In Planck units, $1\ C/m^3 \sim 2\ 10^{-87}$, $r \sim 10^{22}$, $1\ kg/m^3 \sim 2\ 10^{-97}$, so normally $qd^2\ r^2 << md$.<br>We have included electric charge for completeness because it fits everywhere. To compare our results with results from gravitational theories without charge, we set $qd = 0$. This equation enters in equation (29) (Apx D, c83). |

These inequality constraints will be used in section 3.3 to derive the continuum limit.

### *3.3. Translational holonomy / area around a Kerr mass in the continuum limit*

We can restate translational holonomy in terms of the continuum densities md and qd. In the continuum limit, terms that contain a factor of r to a positive power disappear in each expression.

(20) $$\Delta = r^2 + a^2 - 2\ md\ r^4 + qd^2\ r^6$$ (Apx D, e4)

(21) $$c1 = \frac{a}{r}(-1 - md\ r^2 + qd^2\ r^2)$$ (Apx D, e43)

(22) $$c2 = -\frac{\Delta^{1/2}}{r} = -\left(1 - 2\ md\ r^2 + \frac{a^2}{r^2} + q^2\ r^4\right)^{1/2}$$ (Apx D, e43)

(23) $$k1 = \frac{2\ \pi\ a}{\nu}(3\ md\ r^2 - 2\ qd^2\ r^4 + a^2\ md - a^2\ qd^2\ r^2)$$ (Apx D, e44)

(24) $$k2 = (1 - a^2\ md + a^2\ qd^2\ r^2)\ \frac{\Delta^{1/2}}{\nu^2}$$ (Apx D, e45)

(25) $$\nu^2 = (1 - 2\ a^2\ md) + (2\ a^2\ qd^2 - 2\ md - a^2\ md^2)\ r^2 + qd^2\ (1 + 2\ a^2\ md)\ r^4 - a^2\ qd^4\ r^6)$$

Expressed in terms of m, and q, the above result is (Apx D, d43):

$$(26)\qquad v^2 = \left(1 - 2 a^2 \frac{m}{r^3}\right) + \left(2 a^2 \frac{q^2}{r^4} - 2 \frac{m}{r} - a^2 \frac{m^2}{r^4}\right) + \frac{q^2}{r^2}\left(1 + 2 a^2 \frac{m}{r^5}\right) - a^2 \frac{q^4}{r^6}$$

The area of an equatorial loop as seen from infinity is $\pi r^2$. So holonomy per unit area is:

$$(27)\qquad \frac{\text{translational holonomy}}{\text{area}} = \begin{bmatrix} -k1/(\pi r^2) \\ k2 (\cos(2 \pi \nu) - 1) / (\pi r^2) \\ k2 \sin(2 \pi \nu) / (\pi r^2) \\ 0 \end{bmatrix} \qquad \text{(Apx D, d77)}$$

In equation (15), expand k1 and k2, and replace m and q with the densities $md\ r^3$ and $qd\ r^3$ to get (Apx D, d82):

$$(28)\qquad \frac{\text{translational holonomy}}{\text{area}} = \begin{bmatrix} \dfrac{2 \pi a (2\, qd^2 r^4 + a^2 qd^2 r^2 - 3\, md\, r^2 - a^2 md)}{1 - \alpha} \\ \dfrac{(\cos(2 \pi \alpha) - 1)(a^2 qd^2 r^2 - a^2 md + 1)(qd^2 r^6 - 2\, md\, r^4 + r^2 + a^2)^{1/2}}{(1 - \alpha)^2} \\ \dfrac{\sin(2 \pi \alpha)(a^2 qd^2 r^2 - a^2 md + 1)(qd^2 r^6 - 2\, md\, r^4 + r^2 + a^2)^{1/2}}{(1 - \alpha)^2} \\ 0 \end{bmatrix}$$

Apply the restrictions listed in Table 1.

- When we compare our results to well-known results that use m and a but not q, we set charge density $qd = 0$ to simplify the expressions.
- Assume $md\ r^2 << 1$. This implies that $1 - \alpha \sim (1-2\, md\, r^2)^{1/2} \sim (1 - md\, r^2)$.
- Set $\cos(2 \pi\, md\, r^2) = 1$ and $\sin(2 \pi\, md\, r^2) \to 0$.
- Assume $a^2 << r^2$ so that $3 r^2 + a^2 \to 3 r^2$.
- Replace md with $4/3\pi\ md$, where md is the correctly normalized mass density.

This process yields an expression for translational holonomy/area:

$$(29)\qquad \text{Discrete torsion for equatorial loop} \approx \begin{bmatrix} -8 \pi a\, md \\ 0 \\ 0 \\ 0 \end{bmatrix} \qquad \text{(Apx D, d88)}$$

This result is expressed using the orthonormal frame field. When expressed in coordinate frames, the expression for surrogate torsion has the same expression (Apx D, line d89).

This result can be written in a form that exhibits its structure as $-8 \pi$ (spin density) / volume.

$$(30)\qquad \text{surrogate torsion for equatorial loop} \approx \begin{bmatrix} -8 \pi a\, md/(4/3\pi r^3) \\ 0 \\ 0 \\ 0 \end{bmatrix}$$

### *3.4. Comparison of results with Einstein–Cartan theory*

We want holonomy per unit area for a loop of a radius R on the scale of the continuum model; recall that r is the half-distance between Kerr masses; so we can assume that R >> r.

- The holonomy of the large loop is larger by a factor of $(R/r)^2$, compared to that of the small loop around a single small Kerr mass. We can segment the large disk into small regions (for example squares, which introduce some inaccuracy) whose boundaries cancel inside the big disk.
- The area surrounded by the large loop is larger by a factor of $(R/r)^2$, compared to the value for the small loop of radius r. This result is exact for a flat space and approximately true for a continuum model with a sufficiently large radius of background curvature, based on the estimate of the impact of background curvature $R_b$ above.

Therefore, the translational holonomy per unit area of the loop of radius R is equal to that for a single small Kerr mass surrounded by a small loop.

The field equations of EC result from varying the Lagrangian with respect to

- the translational connection coefficients $\Gamma_\mu{}^a$ (not by varying the metric, which results in a symmetric tensor in place of the more general momentum tensor of EC)

(31) $$G_a{}^\mu = 8\,\pi\,K\,P_a{}^\mu$$

- and the rotational connection coefficients $\Gamma_\mu{}^{ab}$

(32) $$ST_{ab}{}^\mu = 8\,\pi\,K\,Spin_{ab}{}^\mu$$ (Apx D, c93)

where G and P are the Einstein tensor and momentum tensor respectively, Spin is the intrinsic a.m. (called spin in quantum theory), and ST is the modified torsion tensor.

Using Theorem 1, Theorem 2, and the calculations in Section 3.4, the torsion and the spin density are related as in the spin–torsion field equation of EC.

The absence of the torsion trace terms in the heuristic result is due to the simplicity of the rotating mass in this calculation. The torsion trace also vanishes for Dirac fields. The torsion trace apparently describes more complex rotational moments than the first-order moments of the Kerr solution or of Dirac fields. The remaining parts of torsion are also not included in this derivation.

### *3.5. Limitations of the continuum derivation*

#### 3.5.1. Inequality constraints

This computation includes inequality constraints (17) an (18) in Table 1 that restrict the size of the loop radius r. These constraints mean that limits $r \to 0$ are limits in classical physics that do not go all the way to zero. Classical physics has numerous situations where taking limits for small volumes of gravitational or electrical field energies generates infinite energy densities.

#### 3.5.2. Background curvature

Background curvature $R_b$ due to the distribution of Kerr masses makes the area of the disk smaller by roughly a factor of $(1 - (r/R_b)^2/6)$, which is negligible when r is small compared to $R_b$.

We could in principle include the background curvature as follows. Start with a GR solution with continuum mass density md and continuum spin density a md. Gradually reduce the density of the continuum mass, while adding a distribution of small Kerr masses that keep the mass and spin densities of the model constant, except for small fluctuations.

### 3.5.3. Continuum limit uses highly symmetric configurations of matter

This derivation applies only to classical arrays of discrete masses for which the densities of mass, momentum, a.m. (and electric charge) converge to constants. We use a configuration with constant densities because it is the most basic case in all continuum field theories.

### 3.5.4. Mathematical rigor

The continuum derivation does not meet the highest standards of mathematical rigor. I can imagine two more rigorous types of convergence argument.

- Establish convergence of the distributions of Kerr masses to EC in Sobolev norm (the preferred norm for defining convergence of approximations to exact solutions of partial differential equations).
- Second-best would be a numerical computation – using adaptive finite element meshes with gradually decreasing error in Sobolev norm – to establish numerical convergence of approximate solutions for a distribution of Kerr masses to an exact EC solution.

An authoritative discussion of the role of rigor in mathematics and mathematical physics can be found in the *Bulletin of the American Mathematical Society* [BAMS, 1993-1994].

## 4. Conclusion and discussion

This section summarized results of the work and discusses possible applications of EC.

This work presents two derivations of EC from GR.

- Part One derives translational holonomy and translational holonomy/area – the discrete version of torsion – around one exterior Kerr solution. This computation is the contravariant equivalent of E. Cartan's covariant differential forms for torsion and curvature. We show that translational holonomy per unit area equals a discrete approximation for mass density. This equality is the discrete version of the spin–torsion field equation of EC.
- Part Two derives classical continuum EC from GR. It constructs a sequence of distributions of exterior Kerr solutions that converges to a classical spin fluid (a non-quantum fluid that conserves 4-momentum and intrinsic a.m.) that has constant densities of mass and intrinsic a.m. The continuum limit yields torsion and the spin–torsion field equation of EC, subject to the inequality restrictions in Table 1, section 3.2.

Key results include:

1) Only GR, EC, and teleparallel gravity satisfy all current empirical tests [Blagojević and Hehl 2013]. EC is a modest extension of GR whereas teleparallel gravity is equivalent to GR.
2) EC is the only extension of GR that has been derived from GR with no additional assumptions except the inequality constraints in section 3.5, and use of a simple, highly symmetric distribution of Kerr masses.
3) EC is a minimal extension of GR that describes exchange of intrinsic and orbital angular momentum. GR is genetically unable to do this because its momentum tensor must be symmetric.
4) EC removes some gravitational singularities that occur in GR in regions of very high spin density: in black holes and Big Bang cosmological models.
5) EC introduce a spin contact force that is strong in regions with high spin density. As far as is known, this contact force is a viable candidate as a cause of cosmic inflation [Fabbri 2018].
6) Derivation of EC from GR strengthens the case for EC and new physics derived from EC.
7) Riemann–Cartan geometry is the minimal extension of Riemannian geometry that gauges rotational and translational affine symmetries.
    - EC deploys E. Cartan's full theory of affine differential geometry: translations are included in the structure group and structure equations, and torsion is a full partner with rotational curvature.

- The translations in the structure group yield momentum as the Noether current of translational symmetry (which GR does imperfectly using Lie translations in the spacetime manifold).

In summary, we now have four main arguments for EC: the gauge theory argument of Sciama and Kibble; the ability of EC to model exchange of classical intrinsic and orbital a.m.; EC (alone except for teleparallel gravity) satisfies all known empirical tests of GR; and a derivation of the elements of discrete and continuum EC from GR, subject to the limitations discussed in section 3.5.

Below are some speculative applications of EC.

- EC may provide a better classical limit for theories of quantum gravity, because treatment of intrinsic angular momentum is more important in quantum theories than in their classical counterparts.
- In EC, the largest force in some Friedmann models of the very early universe is the repulsive force of torsion, so EC may provide a geometric explanation for cosmic inflation.
- Turbulence, vortices, and viscous heating by dynamical friction are topics of current research on dark matter. [Bhatt 2019; Hui 2020; Yang 2020]. If dark matter includes turbulence, then torsion and EC may have a significant role in future theories of dark matter.

## 5. Post-publication note

May 30, 2022: [Kibble 1961] and [Sciama 1962] assumed presence of torsion and/or the Poincaré group. They assembled the current version of EC by identifying torsion with intrinsic a.m., and they demonstrated that the resulting theory models the exchange of intrinsic and orbital a.m. They did not use the basic definition of torsion as the limit of (translational holonomy)/(loop area) that enables this work to derive torsion from GR.

In the following decades, EC continued to be treated as an unproven speculation.

- [Trautman 2006] credits EC for satisfying all empirical tests of GR, but treats EC as an unproven speculation because it has no empirical validation of its own, nor has any work on EC approached a proof or derivation.
- [Blagojević and Hehl 2013] takes essentially the same position about EC as [Trautman 2006].
- In 1989, I asked Yuval Ne'eman whether he could prove EC; he had no answer. I explained Part One of this work and gave him my 1986 paper. Later Ne'eman wrote that "your work on GR and the EC "theory" (quotes indicate that EC is no longer a speculation) is of the highest quality and I have often quoted your results,"
- Between 1960 and 2010, virtually no one who worked primarily on torsion gravity received tenure at a research institution in the USA. The most important work on EC was done in Europe.

Unlike earlier work on EC, this work derives EC from only pseudo-Riemannian geometry and classical GR – without assuming torsion, the Poincaré group or spinors. It uses the deep definition of curvature as limit[area→0] (holonomy)/(loop area) to derive torsion without assuming it. (This is the contravariant version of Cartan's classical definition of curvature as the exterior covariant derivative of the connection form.) Torsion appears in EC exactly where the Cartan structural equations of affine geometry prescribe, which makes EC more credible theories of torsion gravity that contain ad-hoc terms.

Some commentaries on this work are:

- Some physicists maintain that this work is an ad-hoc treatment of EC, on a par with [Kibble 1961], [Sciama 1962]. Therefore it is no more definitive than those works. This viewpoint treats the contravariant version of definition of torsion (that we use to derive torsion) as an ad-hoc assumption. I believe those who hold this view are unfamiliar with the contravariant definition of curvature including torsion.
- Physicists generally assign little credibility to a theory that has no unique empirical evidence. In my opinion, applying this principle to a tight derivation of EC from GR, without any additional assumptions or ad hoc terms, is an overreach of the empiricist viewpoint.

In my opinion, this derivation, and EC's ability to model exchange of intrinsic and orbital a.m. (which occurs in cosmology as well as in microphysics), and the absence of ad-hoc torsion terms establish that EC is, in theory, a necessary extension of GR.

No standard text on gravitation treats EC. There is no published collection of basic EC solutions. EC should have a major role in research on the initial state of the cosmos, cosmic inflation, and black holes.

**Appendix A: Rotational curvature and affine torsion as limits of holonomy**

Many gravitational physicists believe that torsion cannot be derived from Riemannian geometry without an independent assumption. The present work circumvents this claim in two stages: This appendix derives rotational curvature and torsion as limits of rotational and translational holonomy respectively. Appendix G derives rotational holonomy and translational holonomy from a flat manifold.

Greek indices indicate spacetime indices; small Roman letters a, b, c, … represent frame indices; small Roman indices i, j, k, … represent spatial frame indices. Vectors without indices appear in bold underlined font **v**. For a summary of notations, see Appendix F.

A.1. *Development of curves*

T[1]he motive for defining development of curves is isolate the influence of local manifold geometry from the influence of acceleration upon the shape of a curve.

- Ξ is a smooth (pseudo-) Riemannian manifold of dimension n with local coordinates $\xi^\mu$, metric $g_{\mu\nu}$, and metric connection coefficients $\Gamma_{\lambda\nu}{}^{\mu}$ whose covariant differentiation is denoted by D.
- C: [0, 1] → Ξ is a smooth curve with $C(0) = \xi_0$ and tangent vector field $\bar{\mathbf{u}}$.
- A(σ): $T_{C(0)}\Xi \rightarrow T_{C(\sigma)}\Xi$ is an isometry of the tangent space at C(0) with the tangent space at C(σ), by parallel translation along curve C.

**Lemma A.1:** The mapping A(σ) satisfies the differential system

(A-1) $$\frac{dA^\mu{}_\nu}{d\sigma}(\sigma) + u^\lambda(\sigma)\,\Gamma_{\lambda\mu}{}^{\nu}(\sigma)\,A^\mu{}_\nu(\sigma) = 0$$

with initial condition $A^\mu{}_\nu(0) = \text{kronecker_delta}^\mu{}_\nu$.

The index μ refers to a vector basis at C(σ), while the index ν refers to a vector basis at C(0). The proof of Lemma 1 consists of parallel translating the vector along C from C(0) to C(σ).

Define similar structures for another manifold Ξ':

- Ξ' is a flat smooth (pseudo-) Riemannian manifold of the same dimension as Ξ with local coordinates $\xi'^\mu$, metric $g'_{\mu\nu}$, and metric connection coefficients $\Gamma'_{\lambda\nu}{}^{\mu}(\sigma)$ whose covariant differentiation is denoted by D'.
- C': [0, 1] → Ξ' is a smooth curve with $C'(0) = \xi'_0$ and tangent vector field u'.
- A'(σ): $T_{C'(0)}\Xi' \rightarrow T_{C'(\sigma)}\Xi'$ is an isometry of the tangent space at C'(0) with the tangent space at C'(σ), by parallel translation along the curve C'.

Applying Lemma A.1 to the curve on manifold Ξ', the mapping A'(σ) satisfies the differential system

(A-2) $$\frac{dA'^a{}_b}{d\sigma}(\sigma) + u'^c(\sigma)\,\Gamma'_{ca}{}^{b}(\sigma)\,A'^a{}_b(\sigma) = 0$$

with initial condition $A'^\mu{}_\nu(0) = \text{kronecker_delta}'^\mu{}_\nu$.

Choose an isometry of the tangent spaces at the initial points of the two curves, L: $T_\xi\Xi \rightarrow T_{\xi'}\Xi'$.

**Definition:** The development of the curve C in Ξ is the curve C': [0, 1] → Ξ' defined by:

1

(A-3) $$D'_{\underline{u}'}\, \underline{u}' = A'(\sigma)\, L\, [A(\sigma)]^{-1}\, D_u\, \underline{u}$$

with $C'(0) = \xi'_0$ and $\underline{u}'(0) = L\, \underline{u}(0)$.

The development of curve $C$ parallel translates $\underline{u}(\sigma)$ at $C(\sigma)$ back to the point $\xi_0$, then maps it to a vector at $\xi'_0 \in \Xi'$, then parallel translates the vector from $C'(0)$ to $C'(\sigma)$.

A.2. *Definition of curvature and torsion in terms of holonomy*

Rotational holonomy is the linear transformation $g_t$ that maps the starting vector basis onto the ending vector basis that results from parallel translation around the closed loop $C_t$.

Rotational curvature is the rotational holonomy per unit area enclosed by the loop $C_t$, in the limit as the area of the loop approaches zero [Ambrose & Singer, 1953].

(A-4) $$-R(v \wedge w) = \lim_{t \to 0} \frac{g_t}{\text{area}(t)}$$

The developed curves $C'_t$ described above generally do not close. Translational holonomy is defined as the failure-to-close vector needed to close the developed curve, $C'_t(1) - C'_t(0)$.

Affine torsion is the translational holonomy per unit area enclosed by the loop $C'_t$, in the limit as the area of the loop approaches zero.

(A-5) $$-T(v \wedge w) = \lim_{t \to 0} \frac{(C'_t(1) - C'_t(0))}{\text{area}(t)}$$

A.3. *Rotational holonomy in the Schwarzschild solution*

To illustrate the technique calculating curvature as the limit of holonomy/area, we calculate the rotational holonomy of an equatorial circular orbit in the exterior Schwarzschild solution with metric

(A-6) $$ds^2 = -(1 - 2\, m/r)\, dt^2 + \frac{dr^2}{(1 - 2\, m/r)} + r^2\, (d\theta^2 + \sin^2(\theta)\, d\varphi^2)$$

The orbit has $\theta = \pi/2$, so the rotational holonomy is represented as

(A-7) $$2\pi\, [1 - (1 - 2\, m/r)^{1/2}]$$

In the limit $m << r$, the holonomy per unit area enclosed by the loop (using coordinate area, where $r$ is derived from the area of the sphere of radius $r$) becomes approximately

(A-8) $$\frac{2\, m}{r^3} = \frac{8}{3}\pi \frac{m}{(4\,\pi/3)\, r^3} \qquad \text{(assumes } m << r\text{)}$$

The measure of area of the closed loop is that seen by an observer at spatial infinity. This expression is an discrete surrogate for the sectional curvature $R(r, \varphi)$. Three of these terms provide a surrogate for the scalar curvature

(A-9) $$R = R_{ij}{}^{ji} \qquad \text{(note that } R_{0ij0} = 0\text{)}$$

With $m/(4\,\pi/3\, r^3)$ approximating the mass density, we have recovered the field equations of GR that equate the scalar curvature to $8\,\pi$ rho (where rho is the energy density in a fluid model of matter).

## Appendix B: Proof of theorem 1 (translational holonomy of equatorial loops)

B.1. *Holonomic coordinate bases and orthonormal frames*

A basis of the tangent space is called "holonomic" if and only if the Lie brackets of the basis vectorfields are zero; equivalently, if and only if there exist local coordinate systems that yield the basis vectors.

Throughout the entire article, we use the following symbols.

$a := S/m$

$\Delta := r^2 + a^2 - 2\,m\,r + q^2$

$\rho^2 := r^2 + a^2 \cos^2(\theta)$

$A := -2\,m\,r + q^2$

$c := \cos(\theta),\ s := \sin(\theta)$

Two bases will be used for the tangent vectors and covectors on the Kerr manifold. The first is the coordinate basis: $(\partial/\partial t, \partial/\partial r, \partial/\partial\theta, \partial/\partial\varphi)$; with dual basis $(dt, dr, d\theta, d\varphi)$.

The second basis is an orthonormal frame field. (Apx D, d6)

$$e_0 = \frac{1}{\rho\,\Delta^{1/2}} \left[ (r^2 + a^2) \frac{\partial}{\partial t} + a \frac{\partial}{\partial \varphi} \right]$$

$$e_1 = \frac{\Delta^{1/2}}{\rho} \frac{\partial}{\partial r}$$

$$e_2 = \frac{1}{\rho} \frac{\partial}{\partial \theta}$$

$$e_3 = \frac{1}{\rho \sin(\theta)} \left[ a \sin^2(\theta) \frac{\partial}{\partial t} + a \frac{\partial}{\partial \varphi} \right]$$

The dual basis of the orthonormal frame field is (Apx D, c2):

$$e^0 = \frac{\Delta^{1/2}}{\rho} \left[ dt - a \sin^2(\theta)\, d\varphi \right]$$

$$e^1 = \frac{\rho\, dr}{\Delta^{1/2}}$$

$$e^2 = \rho\, d\theta$$

$$e^3 = \frac{\sin(\theta)}{\rho} \left[ (r^2 + a^2)\, d\varphi - a\, dt \right]$$

B.2. *Lie brackets of frame fields*

A basis of the tangent space is called “anholonomic” if and only if the Lie brackets of the basis vector fields are non-zero.

The Lie brackets of the frame fields are specified frame brackets coefficients fb as in the following formula.

$$[e_a, e_b] = fb_{ab}{}^{c}\, e_c \qquad a, b, c = 0,1,2,3$$

The non-vanishing components of the frame brackets are shown in Appendix D, lines e9 – e15.

B.3. *Connection coefficients*

The connection coefficients with three frame field indices are computed first; the non-zero coefficients are displayed in Appendix D, lines e7 – e40.

Connection coefficients frame field indices = $\Gamma_{ab}{}^{c}$

These are used only to compute the connection coefficients with mixed spacetime and fiber indices described next.

The computation of holonomy uses connection coefficients with a spacetime index in the first position (the direction of covariant differentiation), and frame indices in the last two positions (the rotation of frames).

$$\text{Connection coefficients with mixed indices} = \Gamma_{\mu b}{}^{a}$$

where $\mu$ is a coordinate index and $a$ and $b$ are frame field indices. These connection coefficients are displayed in Appendix D in the five matrices after line c42.

B.4. *Development of equatorial loop into Minkowski space*

The starting point and initial tangent of the equatorial loop used in Theorem 1 are given below.

$$\text{(B-1)} \qquad \underset{\text{(Kerr coords)}}{C^{\mu}(0)} = \begin{vmatrix} 0 \\ r \\ 0 \\ 0 \end{vmatrix} \qquad \underset{\text{(Kerr coords)}}{d\,C^{\mu}(0)/d\sigma} = \begin{vmatrix} 0 \\ 0 \\ 0 \\ 2\pi \end{vmatrix} \qquad \underset{\text{(frames)}}{d\,C^{a}(0)/d\sigma} = \begin{vmatrix} -\dfrac{2\pi a \Delta^{1/2}}{r} \\ 0 \\ 0 \\ \dfrac{2\pi (r^2+a^2)}{r} \end{vmatrix}$$

The acceleration vector in coordinate and orthonormal frames (Apx D, d50 – d52):

$$\text{(B-2)} \qquad \underset{\text{(Kerr cords)}}{\mathrm{Accel}^{\mu}(\sigma)} = \begin{vmatrix} -\dfrac{2\pi a \Delta^{1/2}}{r} \\ 0 \\ 0 \\ \dfrac{2\pi a(r^2+a^2)}{r} \end{vmatrix} \qquad \underset{\text{(frames)}}{\mathrm{Accel}^{a}(\sigma)} = \begin{vmatrix} 0 \\ -\dfrac{4\pi^2 \Delta^{1/2}(r^4 - a^2 m r + a^2 q^2)}{r^4} \\ 0 \\ 0 \end{vmatrix} = \begin{vmatrix} 0 \\ -4\pi^2\, k2\, \nu^2 \\ 0 \\ 0 \end{vmatrix}$$

We want to use the mapping $A^{a}{}_{b}(\sigma)$ that parallel translates vectors at the start of the equatorial loop to any point on the loop with parameter value $\sigma$. The indices on $A^{a}{}_{b}$ are orthonormal frame indices.

Lemma B.1 states that $A(\sigma)$ satisfies the differential system

$$dA^{a}{}_{b}(\sigma)/d\sigma + u^{\lambda}(\sigma)\, \Gamma_{\lambda c}{}^{a}(\sigma)\, A^{c}{}_{b}(\sigma) = 0$$

In Kerr coordinates, on the specified equatorial loop, this equation is:

$$\text{(B-3)} \qquad dA^{a}{}_{b}(\sigma)/d\sigma + 2\pi\, \Gamma_{\varphi c}{}^{a}(\sigma)\, A^{c}{}_{b}(\sigma) = 0$$

The matrix $A(\sigma)$ below integrates parallel translation for the chosen equatorial loop satisfies this equation with initial condition $A(0) = \text{identity}$. (Apx D, two matrices displayed after line c55)

Notation: $Z := -r + q^2/r - m \qquad cc := \cos(2\pi\nu\sigma), \quad ss := \sin(2\pi\nu\sigma)$

$$A^a{}_b(\sigma) = \begin{bmatrix} \frac{a Z^2 (1-cc)+1}{\nu^2 r^4} & -\frac{a Z\, ss}{\nu r^2} & 0 & -\frac{a \Delta^{1/2} Z (1\text{-}cc)}{\nu^2 r^3} \\ -\frac{a Z\, ss}{\nu r^2} & \frac{(a^2 Z^2 - \Delta r^2)(1-cc)+1}{\nu^2 r^4} & 0 & \frac{\Delta^{1/2}\, ss}{\nu r} \\ 0 & 0 & 1 & 0 \\ \frac{a (1-cc) \Delta^{1/2}}{\nu^2 r^3} & -\frac{ss\, \Delta^{1/2}}{\nu r} & 0 & 1 - \frac{(1-cc)\Delta}{\nu^2 r^2} \end{bmatrix} \tag{B-4}$$

Pull vector **Accel**(σ) at C(σ) back to the starting point C(0) by multiplying by $A^{-1}(\sigma)$. (Apx D, d58)

$$\textbf{Accel}(\sigma) = A^{-1}(\sigma) \begin{bmatrix} \frac{4 \pi^2 a\, k2\, \nu\, (r^2 + m r - q^2) \sin(2 \pi \nu \sigma)}{r^3} \\ -\frac{4 \pi^2 k2 \left(r^4 (r^2 - 2 m r + q^2) - a^2 (m r - q^2)(2 r^2 + m r - q^2)\right) \cos(2 \pi \nu \sigma)}{r^6} \\ 0 \\ -\frac{4 \pi^2 k2\, \nu\, \Delta^{1/2}\, \sigma\text{in}(2 \pi \nu \sigma)}{r} \end{bmatrix} \tag{B-5}$$

The next step is to choose a fixed isometry $L^\mu{}_\nu$ between the tangent vectors at initial points C(0) in spacetime Ξ and C'(0) in the development space Ξ'. We have chosen the starting point and initial direction of the developed curve so that the solutions in Minkowski coordinates reflect the rotational symmetry of the problem. (Apx D, d59)

$$L^a{}_\nu = \begin{bmatrix} -\frac{\Delta^{1/2}}{\nu r} & 0 & 0 & \frac{a Z}{\nu r^2} \\ 0 & 1 & 0 & 0 \\ -\frac{a Z}{\nu r^2} & 0 & 0 & \frac{\Delta^{1/2}}{\nu r} \\ 0 & 0 & 1 & 0 \end{bmatrix} \tag{B-6}$$

Since we use Minkowski coordinates (t', x', y', z') on the flat space Ξ', parallel translation in local coordinates is trivial. The acceleration of curve C(σ) is mapped from C(σ) to C(0) to C'(0) to C'(σ) via that mapping $A'(\sigma)\, L\, [A(\sigma)]^{-1}$, is The acceleration of the developed curve in Minkowski space is (Apx D, d61):

$$\text{Accel}'(\sigma) = \begin{vmatrix} 0 \\ 4 \pi^2 k2\, \nu^2 \cos(2 \pi \nu \sigma) \\ 4 \pi^2 k2\, \nu^2 \sin(2 \pi \nu \sigma) \\ 0 \end{vmatrix} \tag{B-7}$$

The initial condition dC'(0)/dσ is determined by the orthogonal map L, and the choice of the initial point C'(0) of the developed curve is arbitrary. We choose both of these to express the rotational symmetry.

The initial conditions for development of the equatorial loop in Minkowski space (coordinates (t, x, y, z) are (Apx D, d61 and d64):

$$\text{(B-8)}\qquad C'(0) = \begin{vmatrix} 0 \\ k2 \\ 0 \\ 0 \end{vmatrix} \qquad d\,C'(0)/d\sigma = \begin{vmatrix} -\,k1 \\ 0 \\ 2\,\pi\,k2\,\nu \\ 0 \end{vmatrix}$$

The development of the equatorial loop into Minkowski space is (Apx D, d68):

$$\text{(B-9)}\qquad C'(\sigma) = \begin{vmatrix} -\,k1\,\sigma \\ k2\cos(2\,\pi\,\nu\,\sigma) \\ k2\sin(2\,\pi\,\nu\,\sigma) \\ 0 \end{vmatrix}$$

B.5. *Translational holonomy of an equatorial loop developed into Minkowski space*

The translational holonomy (“failure-to-close vector”) of the equatorial loop in Minkowski space is:

$$\text{(B-10)}\qquad C'(1) - C'(0) = \begin{vmatrix} -\,k1 \\ k2\,(\cos(2\,\pi\,\nu) - 1) \\ k2\sin(2\,\pi\,\nu)) \\ 0 \end{vmatrix}$$

This completes the proof of Theorem 1.

## Appendix C: Proof of theorem 2 (translational holonomy of polar loops)

For development of a curve in the $\theta$ direction extending from $\theta_0$ to $\theta_1$, the development matrix $A$ is:

$$\text{(C-1)}\qquad A^a{}_b = \begin{vmatrix} \cosh(F) & 0 & 0 & \sinh(F) \\ 0 & \cos(G) & -\sin(G) & 0 \\ 0 & \sin(G) & \cos(G) & 0 \\ \sinh(F) & 0 & 0 & \cosh(F) \end{vmatrix}$$

where

$$\text{(C-2)}\qquad F = \left| -\frac{\Delta}{r^2 + a^2} \tanh^{-1} \left[ \frac{a\sin(\theta)}{(r^2 + a^2)^{½}} \right] \right|_{\theta_0}^{\theta_1}$$

$$\text{(C-3)}\qquad G = \left| -\frac{\Delta}{r^2 + a^2} \tanh^{-1} \left[ \frac{r\tan(\theta)}{(r^2 + a^2)^{½}} \right] \right|_{\theta_0}^{\theta_1}$$

The tensor indices $a$ and $b$ refer to frame field directions. When $\theta_0$ and $\theta_1$ are multiples of $\pi$, both $F$ and $G$ vanish, so the matrix $A$ is the identity. Therefore the development of a closed loop in the $\theta$ direction is closed, so the translational holonomy is zero.

**Appendix D: Executable computer algebra derivation of results**

This executable computer algebra script provides traceable derivations of formulas in the main article.

Computer algebra system Macsyma 2.4.1a was used to verify most of the computations in this work [Macsyma 1996]. This appendix is the script of that validation. As above, we use lower case Greek letters for coordinate indices and lower-case Roman letters for fiber indices.

**Outline of the Macsyma computer algebra script**

Part I: Translational holonomy and torsion around one Kerr mass

1. Geometry of the Kerr-Newman metric
2. Connection coefficients
3. Translational holonomy and surrogate torsion around and equatorial loop

Part II: Continuum limit of a distribution of discrete rotating masses

4. Holonomy and torsion in the continuum limit
5. Spin density and the spin-torsion relationship

The computations om Parts I and II below illustrate how an orthonormal frame field can exploit the symmetry of the metric to greatly simplifies results.

**Conversions between conventional tensor notations and Macsyma notations**

| Description | Conventional notation | Macsyma notation | Relationships |
|---|---|---|---|
| Covariant metric in coordinate basis | $g_{\mu\nu}$ | lg | |
| Contravariant metric in coordinate basis | $g^{\mu\nu}$ | ug | $g^{\mu\nu}$ = invert($g_{\mu\nu}$) |
| Frame field | $fr_a{}^{\mu}$ | fr | |
| Inverse frame field | $fri_{\mu}{}^{a}$ | fri | $fri_{\mu}{}^{a}$ = invert($fr_a{}^{\mu}$) |
| Covariant metric in frame basis | $g_{ab}$ | lfg | $g_{ab} = fr_a{}^{\mu}\, g_{\mu\nu}\, fr_b{}^{\nu}$ |
| Contravariant metric in frame basis | $g^{ab}$ | ufg | $g^{ab} = fri_{\mu}{}^{a}\, g^{\mu\nu}\, fri_{\nu}{}^{b}$ |
| Christoffel symbols in frame basis | $mcs_{cb}{}^{a}$ | mcs | |
| Christoffel symbols in mixed coordinate-frame basis | $mcs_{\mu b}{}^{a}$ | mcs_cff | $mcs_{\mu b}{}^{a} = fri_{\mu}{}^{c}\, mcs_{cb}{}^{a}$ |

Kerr coordinates: $t$ = time, $r$ = radius, $\theta$ = azimuthal angle (latitude), $\varphi$ = rotational angle (longitude)

# Part One: Translational Holonomy of an Equatorial Loop Around a Charged Rotating Mass

This computer algebra script was created with Macsyma 2.4.1.a using the component tensor package.

## 1. Geometry of the Kerr Newman Metric

### 1.1 Metric and Frame Fields

Index conventions:

* A lower-case Greek tensor index indicates a spacetime (coordinate) index.
* A lower-case Roman index indicates a frame index.
* In commands (c-lines), "@" before a tensor index indicates that the index is contravariant (raised).

Enter coordinates [t, r, θ, φ], turn on frame fields, and set lowered frame metric $LFG_{ab}$ = Minkowski metric.

**(c1)**
```
(init_ctensor(), ratfac : true, dim : 4, ct_coords : [t, r, theta, phi], assume(m > 0),
  cframe_flag : true, lfg : diag_matrix(-1,1,1,1) )
```

**(d1)**

$$\begin{bmatrix} -1 & 0 & 0 & 0 \\ 0 & 1 & 0 & 0 \\ 0 & 0 & 1 & 0 \\ 0 & 0 & 0 & 1 \end{bmatrix}$$

Enter inverse frame field $FRI_\mu{}^i$ that defines the Kerr Newman metric.

If $d/dx^m$ are coordinate basis fields and $E_a$ are frame fields, then $E_a = d/dx^\mu\, FR_a{}^\mu$ .

Think of FRI as the “square root of the metric tensor” since the lowered metric $LG_{mn} = FRI_\mu{}^a\, LFG_{ab}\, FRI_\nu{}^b$.

**(c2)**
```
fri : matrix(sqrt(\delta)*[1, 0, 0, -a*sin(theta)^2]/rho, [0, rho/sqrt(\delta), 0, 0],
   [0, 0, rho, 0], sin(theta)*[-a, 0, 0, r^2+a^2] / rho)
```

**(d2)**

$$\begin{bmatrix} \frac{\sqrt{\Delta}}{\rho} & 0 & 0 & -\frac{a \sin^2(\theta)\sqrt{\Delta}}{\rho} \\ 0 & \frac{\rho}{\sqrt{\Delta}} & 0 & 0 \\ 0 & 0 & \rho & 0 \\ -\frac{a\sin(\theta)}{\rho} & 0 & 0 & \frac{(r^2+a^2)\sin(\theta)}{\rho} \end{bmatrix}$$

Define derivatives of Δ and ρ with respect to the coordinates r and θ.

**(c3)**
```
(gradef(\delta, r, 2*(r-m)), gradef(rho, r, r/rho),
 gradef(rho, theta, -a^2*sin(theta)*cos(theta)/rho), assume(rho>0))$
```

Define some transformation rules for expressions that appear frequently.

(c4)
```
(defrule(delta_def, r^2-2*m*r+a^2+q^2, \delta),defrule(delta_expand, \delta, r^2-2*m*r+a^2+q^2),
 defrule(rho2_def1, r^2+(a*cos(theta))^2, rho^2), defrule(rho2_def2, a^2*sin(theta)^2-r^2-a^2, -rho^2),
 defrule(rho2_expand, rho^2, r^2+(a*cos(theta))^2), defrule(rho_expand, rho, sqrt(r^2+(a*cos(theta))^2)),
 defrule(r2_a2, a^2*sin(theta)^2+rho^2, r^2+a^2), defrule(mr_expand, r^2+a^2-\delta, 2*m*r-q^2),
disprule(delta_expand, rho2_expand))$
```

(e4)
$$\text{delta_expand}(\text{defrule}) : \Delta \to r^2 - 2\,m\,r + q^2 + a^2$$

(e5)
$$\text{rho2_expand}(\text{defrule}) : \rho^2 \to a^2 \cos^2(\theta) + r^2$$

Compute and display frame field $FR_a{}^{\mu}$, lowered metric $LG_{\mu\nu}$. Compute but do not display raised metric $UG^{\mu\nu}$ .

(c6) `(cmetric(), fr : apply1(factor(fr), rho2_def2))`

(d6)
$$\begin{bmatrix} \frac{r^2+a^2}{\rho\sqrt{\Delta}} & 0 & 0 & \frac{a\sin(\theta)}{\rho} \\ 0 & \frac{\sqrt{\Delta}}{\rho} & 0 & 0 \\ 0 & 0 & \frac{1}{\rho} & 0 \\ \frac{a}{\rho\sqrt{\Delta}} & 0 & 0 & \frac{1}{\rho\sin(\theta)} \end{bmatrix}$$

Re-express the metric tensor $LG_{\mu\nu}$ using transformation rule MR_EXPAND.

(c7) `lg : apply1(lg, mr_expand)`

(d7)
$$\begin{bmatrix} -\frac{\Delta - a^2\sin^2(\theta)}{\rho^2} & 0 & 0 & \frac{a\sin^2(\theta)\left(\Delta - r^2 - a^2\right)}{\rho^2} \\ 0 & \frac{\rho^2}{\Delta} & 0 & 0 \\ 0 & 0 & \rho^2 & 0 \\ \frac{a\sin^2(\theta)\left(\Delta - r^2 - a^2\right)}{\rho^2} & 0 & 0 & -\frac{\sin^2(\theta)\left(a^2\sin^2(\theta)\,\Delta - r^4 - 2\,a^2\,r^2 - a^4\right)}{\rho^2} \end{bmatrix}$$

## 1.2 Frame Brackets

The Lie brackets of the frame fields were computed by command CMETRIC above.

Apply simplification rules to the frame brackets.

(c8)
```
(for i thru dim do for j thru dim do for k thru dim do fb[i,j,k] :
  apply1(trigsimp(apply1(fb[i,j,k],delta_expand,rho2_expand)), rho2_def1,delta_def),
 cdisplay(fb) )$
```

(e9)
$$fb_{1,2,4} = \frac{2\,a\,r\sin(\theta)}{\rho^3}$$

(e10)
$$fb_{1,3,1} = -\frac{a^2\cos(\theta)\sin(\theta)}{\rho^3}$$

**(e11)**
$$fb_{2,3,2} = -\frac{a^2 \cos(\theta)\sin(\theta)}{\rho^3}$$

**(e12)**
$$fb_{2,3,3} = -\frac{r\sqrt{\Delta}}{\rho^3}$$

**(e13)**
$$fb_{2,4,4} = -\frac{r\sqrt{\Delta}}{\rho^3}$$

**(e14)**
$$fb_{3,4,1} = \frac{2\,a\cos(\theta)\sqrt{\Delta}}{\rho^3}$$

**(e15)**
$$fb_{3,4,4} = -\frac{\left(r^2 + a^2\right)\cos(\theta)}{\rho^3 \sin(\theta)}$$

# 2. Connection Coefficients

## 2.1 Connection Coefficients with Frame Fields

Connection coefficieints $MCS_k{}^a{}_b$ with one index up and two down, using all frame fields

Index k is the direction of covariant differentiation.

**(c16)** christof(false)$

Apply simplification rules to the connection coeffiicents.

**(c17)**
```
(for i thru dim do for j thru dim do for k thru dim do
  mcs[i,j,k] : apply1(trigsimp(apply1(mcs[i,j,k],delta_expand,rho2_expand)),rho2_def1,delta_def),
 cdisplay(mcs) )$
```

**(e17)**
$$mcs_{1,1,2} = \frac{\left(a^2 r - a^2 m\right)\cos^2(\theta) + m\,r^2 + \left(-q^2 - a^2\right) r}{\rho^3 \sqrt{\Delta}}$$

**(e18)**
$$mcs_{1,1,3} = -\frac{a^2 \cos(\theta)\sin(\theta)}{\rho^3}$$

**(e19)**
$$mcs_{1,2,1} = \frac{\left(a^2 r - a^2 m\right)\cos^2(\theta) + m\,r^2 + \left(-q^2 - a^2\right) r}{\rho^3 \sqrt{\Delta}}$$

**(e20)**
$$mcs_{1,2,4} = \frac{a\,r\sin(\theta)}{\rho^3}$$

**(e21)**
$$mcs_{1,3,1} = -\frac{a^2 \cos(\theta)\sin(\theta)}{\rho^3}$$

**(e22)**
$$mcs_{1,3,4} = \frac{a\cos(\theta)\sqrt{\Delta}}{\rho^3}$$

**(e23)**

$$mcs_{1,4,2} = -\frac{a\,r\,\sin(\theta)}{\rho^3}$$

**(e24)**

$$mcs_{1,4,3} = -\frac{a\cos(\theta)\sqrt{\Delta}}{\rho^3}$$

**(e25)**

$$mcs_{2,1,4} = -\frac{a\,r\,\sin(\theta)}{\rho^3}$$

**(e26)**

$$mcs_{2,2,3} = \frac{a^2\cos(\theta)\sin(\theta)}{\rho^3}$$

**(e27)**

$$mcs_{2,3,2} = -\frac{a^2\cos(\theta)\sin(\theta)}{\rho^3}$$

**(e28)**

$$mcs_{2,4,1} = -\frac{a\,r\,\sin(\theta)}{\rho^3}$$

**(e29)**

$$mcs_{3,1,4} = \frac{a\cos(\theta)\sqrt{\Delta}}{\rho^3}$$

**(e30)**

$$mcs_{3,2,3} = \frac{r\sqrt{\Delta}}{\rho^3}$$

**(e31)**

$$mcs_{3,3,2} = -\frac{r\sqrt{\Delta}}{\rho^3}$$

**(e32)**

$$mcs_{3,4,1} = \frac{a\cos(\theta)\sqrt{\Delta}}{\rho^3}$$

**(e33)**

$$mcs_{4,1,2} = -\frac{a\,r\,\sin(\theta)}{\rho^3}$$

**(e34)**

$$mcs_{4,1,3} = -\frac{a\cos(\theta)\sqrt{\Delta}}{\rho^3}$$

**(e35)**

$$mcs_{4,2,1} = -\frac{a\,r\,\sin(\theta)}{\rho^3}$$

**(e36)**

$$mcs_{4,2,4} = \frac{r\sqrt{\Delta}}{\rho^3}$$

**(e37)**

$$mcs_{4,3,1} = -\frac{a\cos(\theta)\sqrt{\Delta}}{\rho^3}$$

**(e38)**

$$mcs_{4,3,4} = \frac{\left(r^2+a^2\right)\cos(\theta)}{\rho^3\sin(\theta)}$$

**(e39)**
$$mcs_{4,\,4,\,3} = -\frac{\left(r^2 + a^2\right)\cos(\theta)}{\rho^3 \sin(\theta)}$$

**(e40)**
$$mcs_{4,\,4,\,3} = -\frac{\left(r^2 + a^2\right)\cos(\theta)}{\rho^3 \sin(\theta)}$$

## 2.2 Connection Coefficients with One Coordinate Index

Compute conection coefficients $MCS_{\mu}{}^{a}{}_{b}$ in which the first index is a coordinate index (in the direction of covariant differentiation) and the second and third indices are frame indices (a rotation in fiber space). Simplify.

**(c41)**
```
(array(mcs_cff,4,4,4),
 for b thru 4 do for c thru 4 do for i thru 4 do
  mcs_cff[i,b,c]: apply1(factor(sum(fri[a,i]*mcs[a,b,c],a,1,4)), rho2_def2,trigsimp, mr_expand) )$
```

Express these mixed coordinate / frame-field connection coefficients as matrices.

**(c42)**
```
(array(mcs_cffmat,4),  for i thru 4 do (mcs_cffmat[i] : genmatrix( lambda([b,c],mcs_cff[i,c,b]), 4,4),
 if i <4 then (disp('mcs_cffmat[i]), disp(mcs_cffmat[i]))
  else (disp(['mcs_cffmat[i], "first two columns"], submat(mcs_cffmat[i], [1,2,3,4], [1,2])),
   disp(['mcs_cffmat[i], "last two columns"], submat(mcs_cffmat[i], [1,2,3,4], [3,4])))) )$
```

$$mcs_cffmat_1$$

$$\begin{bmatrix} 0 & -\frac{a^2 m \cos^2(\theta) - m r^2 + q^2 r}{\rho^4} & 0 & 0 \\ -\frac{a^2 m \cos^2(\theta) - m r^2 + q^2 r}{\rho^4} & 0 & 0 & 0 \\ 0 & 0 & 0 & -\frac{a \cos(\theta)\left(\Delta - r^2 - a^2\right)}{\rho^4} \\ 0 & 0 & \frac{a \cos(\theta)\left(\Delta - r^2 - a^2\right)}{\rho^4} & 0 \end{bmatrix}$$

$$mcs_cffmat_2$$

$$\begin{bmatrix} 0 & 0 & 0 & -\frac{a\, r \sin(\theta)}{\rho^2 \sqrt{\Delta}} \\ 0 & 0 & -\frac{a^2 \cos(\theta)\sin(\theta)}{\rho^2 \sqrt{\Delta}} & 0 \\ 0 & \frac{a^2 \cos(\theta)\sin(\theta)}{\rho^2 \sqrt{\Delta}} & 0 & 0 \\ -\frac{a\, r \sin(\theta)}{\rho^2 \sqrt{\Delta}} & 0 & 0 & 0 \end{bmatrix}$$

$$\text{mcs_cffmat}_3$$

$$\begin{bmatrix} 0 & 0 & 0 & \frac{a\cos(\theta)\sqrt{\Delta}}{\rho^2} \\ 0 & 0 & -\frac{r\sqrt{\Delta}}{\rho^2} & 0 \\ 0 & \frac{r\sqrt{\Delta}}{\rho^2} & 0 & 0 \\ \frac{a\cos(\theta)\sqrt{\Delta}}{\rho^2} & 0 & 0 & 0 \end{bmatrix}$$

$$[\text{mcs_cffmat}_4\text{, first two columns}]$$

$$\begin{bmatrix} 0 & -\frac{a\left(a^2 (r-m)\cos^2(\theta)+r^3+m\,r^2-q^2\,r\right)\sin^2(\theta)}{\rho^4} \\ -\frac{a\left(a^2 (r-m)\cos^2(\theta)+r^3+m\,r^2-q^2\,r\right)\sin^2(\theta)}{\rho^4} & 0 \\ -\frac{a\cos(\theta)\sin(\theta)\sqrt{\Delta}}{\rho^2} & 0 \\ 0 & \frac{r\sin(\theta)\sqrt{\Delta}}{\rho^2} \end{bmatrix}$$

$$[\text{mcs_cffmat}_4\text{, last two columns}]$$

$$\begin{bmatrix} -\frac{a\cos(\theta)\sin(\theta)\sqrt{\Delta}}{\rho^2} & 0 \\ 0 & -\frac{r\sin(\theta)\sqrt{\Delta}}{\rho^2} \\ 0 & \frac{\cos(\theta)\left(a^2\sin^2(\theta)\,\Delta-r^4-2\,a^2\,r^2-a^4\right)}{\rho^4} \\ -\frac{\cos(\theta)\left(a^2\sin^2(\theta)\,\Delta-r^4-2\,a^2\,r^2-a^4\right)}{\rho^4} & 0 \end{bmatrix}$$

## 3. Translational Holonomy Around an Equatorial Loop

The frequency ν is defined so the curve forms a closed loop surrounding a Kerr mass in spacetime.

**(c43)**
```
(c_1 : a/r^2*(-r-m+q^2/r), c_2 : - sqrt(\delta)/r,
 nu_exp: sqrt(map('factor, multthru(ratsimp(apply1( c_2^2 - c_1^2, delta_expand))))),
 apply(defrule, ['nu_expand, nu, nu_exp]), disprule(nu_expand))
```

**(d43)**

$$\text{nu_expand}\,(\text{defrule}) : \nu \to \sqrt{-\frac{2\,m}{r}+\frac{q^2}{r^2}-\frac{2\,a^2\,m}{r^3}+\frac{a^2\left(2\,q^2-m^2\right)}{r^4}+\frac{2\,a^2\,m\,q^2}{r^5}-\frac{a^2\,q^4}{r^6}+1}$$

Define k1 and k2, which are used in the solutions for translational holonomy (the displacement vector) in the timelike and equatorial directions respectively.

**(c44)**
```
(defrule(k1_def, (3*m*r^3+a^2*m*r-2*q^2*r^2-a^2*q^2), k1*r^4*nu/(2*%pi*a)),
 defrule(k2_def, (r^4-a^2*m*r+a^2*q^2), k2*r^4*nu^2/sqrt(\delta)),
 defrule(k1_expand, k1, 2*%pi*a*(3*m*r^3+a^2*m*r-2*q^2*r^2-a^2*q^2)/(r^4*nu)),
 defrule(k2_expand, k2, sqrt(\delta)*(r^4-a^2*m*r+a^2*q^2)/(r^4*nu^2)),
 disprule(k1_def, k2_def, k1_expand, k2_expand) )$
```

**(e44)**

$$\text{k1_def}\,(\text{defrule}) : 3\,m\,r^3-2\,q^2\,r^2+a^2\,m\,r-a^2\,q^2 \to \frac{k1\,\nu\,r^4}{2\,\pi\,a}$$

**(e45)**

$$\text{k2_def}\,(\text{defrule}) : r^4-a^2\,m\,r+a^2\,q^2 \to \frac{k2\,\nu^2\,r^4}{\sqrt{\Delta}}$$

**(e46)**

$$\text{k1_expand}\,(\text{defrule}) : k1 \to \frac{2\,\pi\,a\left(3\,m\,r^3-2\,q^2\,r^2+a^2\,m\,r-a^2\,q^2\right)}{\nu\,r^4}$$

**(e47)**

$$\text{k2_expand}\,(\text{defrule}) : k2 \to \frac{\left(r^4-a^2\,m\,r+a^2\,q^2\right)\sqrt{\Delta}}{\nu^2\,r^4}$$

### 3.1 The Equatorial Loop

Define an equatorial loop at radius r around which to compute holonomy. Restrict analyisis to region where r >> m.

**(c48)**
```
(assume(r > m), define(equator(sigma), [0 ; r ; %pi/2; 2*%pi*sigma]))
```

**(d48)**

$$\text{equator}(\sigma) := \begin{bmatrix} 0 \\ r \\ \frac{\pi}{2} \\ 2\,\pi\,\sigma \end{bmatrix}$$

Compute the tangent vector to the path in the coordinate frame.

**(c49)**
```
ucoord : diff(equator(sigma), sigma)
```

**(d49)**

$$\begin{bmatrix} 0 \\ 0 \\ 0 \\ 2\,\pi \end{bmatrix}$$

Express the tangent vector with orthonormal frames. $FRI_{\mu}{}^{a}$ is the inverse frame field expressed as a matrix.

**(c50)** uframe : subst([rho=sqrt(r^2+a^2*cos(theta)^2), theta=%pi/2], fri) . ucoord

**(d50)**

$$\begin{bmatrix} -\dfrac{2\,\pi\,a\,\sqrt{\Delta}}{r} \\ 0 \\ 0 \\ \dfrac{2\,\pi\,\left(r^2+a^2\right)}{r} \end{bmatrix}$$

Compute the acceleration vector with orthonormal frames.

**(c51)** (sum(ucoord[k]* (mcs_cffmat[k] . uframe), k,1,4),
covdiffuuframe1 : ratsimp(subst([rho=sqrt(r^2+a^2*cos(theta)^2),theta=%pi/2], %%), m,q,\delta) )

**(d51)**

$$\begin{bmatrix} 0 \\ -\dfrac{4\,\pi^2\,\left(r^4 - a^2\,m\,r + a^2\,q^2\right)\sqrt{\Delta}}{r^4} \\ 0 \\ 0 \end{bmatrix}$$

Express the acceleration in terms of k2. Define frequency ν so the loop returns to its starting point when σ = 1.

**(c52)** covdiffuuframe2 : apply1( covdiffuuframe1, k2_def)

**(d52)**

$$\begin{bmatrix} 0 \\ -4\,\pi^2\,k2\,\nu^2 \\ 0 \\ 0 \end{bmatrix}$$

### 3.2 Matrix for Parallel Translation along an Equatorial Curve

Matrix A(σ) parallel translates vectors along the equatorial curve from equator(0) to equator(σ).

It satisfies the matrix differential equation below, which is expressed in orthonormal frames.
The initial condition is A(0) = identity.

**(c53)** block([doallmxops:false, domxmxops:false],
'diff(genmatrix(a,4),sigma) + u4 . genmatrix('mcs_cff4,4) . genmatrix(a,4) = 0)

**(d53)**

$$u4\,.\begin{bmatrix} mcs_cff4_{1,1} & mcs_cff4_{1,2} & mcs_cff4_{1,3} & mcs_cff4_{1,4} \\ mcs_cff4_{2,1} & mcs_cff4_{2,2} & mcs_cff4_{2,3} & mcs_cff4_{2,4} \\ mcs_cff4_{3,1} & mcs_cff4_{3,2} & mcs_cff4_{3,3} & mcs_cff4_{3,4} \\ mcs_cff4_{4,1} & mcs_cff4_{4,2} & mcs_cff4_{4,3} & mcs_cff4_{4,4} \end{bmatrix} . \begin{bmatrix} a_{1,1} & a_{1,2} & a_{1,3} & a_{1,4} \\ a_{2,1} & a_{2,2} & a_{2,3} & a_{2,4} \\ a_{3,1} & a_{3,2} & a_{3,3} & a_{3,4} \\ a_{4,1} & a_{4,2} & a_{4,3} & a_{4,4} \end{bmatrix} + \begin{bmatrix} a_{1,1} & a_{1,2} & a_{1,3} & a_{1,4} \\ a_{2,1} & a_{2,2} & a_{2,3} & a_{2,4} \\ a_{3,1} & a_{3,2} & a_{3,3} & a_{3,4} \\ a_{4,1} & a_{4,2} & a_{4,3} & a_{4,4} \end{bmatrix}_{\sigma} = 0$$

Construct the solution matrix Amat.

**(c54)** mat : [0, c_1, 0,0 ; c_1, 0,0, c_2 ; 0,0,0,0 ; 0, -c_2, 0,0]

**(d54)**

$$\begin{bmatrix} 0 & \dfrac{a\left(-r+\dfrac{q^2}{r}-m\right)}{r^2} & 0 & 0 \\ \dfrac{a\left(-r+\dfrac{q^2}{r}-m\right)}{r^2} & 0 & 0 & -\dfrac{\sqrt{\Delta}}{r} \\ 0 & 0 & 0 & 0 \\ 0 & \dfrac{\sqrt{\Delta}}{r} & 0 & 0 \end{bmatrix}$$

Display the solution matrix Amat in two parts: colmumns 1,2 then columns 3,4.

**(c55)**
```
(amat(sigma) := ''(ident(4) + (1-cos(2*%pi*nu*sigma))*mat^^2/nu^2-sin(2*%pi*nu*sigma)*mat/nu),
disp("amat(sigma), columns 1 and 2", submat(amat(sigma), [1,2,3,4], [1,2])),
disp("amat(sigma), columns 3 and 4", submat(amat(sigma), [1,2,3,4], [3,4])) )$
```

amat(sigma), columns 1 and 2

$$\begin{bmatrix} \dfrac{a^2\left(-r+\dfrac{q^2}{r}-m\right)^2\left(1-\cos(2\pi\nu\sigma)\right)}{\nu^2 r^4}+1 & -\dfrac{a\left(-r+\dfrac{q^2}{r}-m\right)\sin(2\pi\nu\sigma)}{\nu r^2} \\ -\dfrac{a\left(-r+\dfrac{q^2}{r}-m\right)\sin(2\pi\nu\sigma)}{\nu r^2} & \dfrac{\left(1-\cos(2\pi\nu\sigma)\right)\left(\dfrac{a^2\left(-r+\dfrac{q^2}{r}-m\right)^2}{r^4}-\dfrac{\Delta}{r^2}\right)}{\nu^2}+1 \\ 0 & 0 \\ \dfrac{a\left(-r+\dfrac{q^2}{r}-m\right)\left(1-\cos(2\pi\nu\sigma)\right)\sqrt{\Delta}}{\nu^2 r^3} & -\dfrac{\sin(2\pi\nu\sigma)\sqrt{\Delta}}{\nu r} \end{bmatrix}$$

amat(sigma), columns 3 and 4

$$\begin{bmatrix} 0 & -\dfrac{a\left(-r+\dfrac{q^2}{r}-m\right)\left(1-\cos(2\pi\nu\sigma)\right)\sqrt{\Delta}}{\nu^2 r^3} \\ 0 & \dfrac{\sin(2\pi\nu\sigma)\sqrt{\Delta}}{\nu r} \\ 1 & 0 \\ 0 & 1-\dfrac{\left(1-\cos(2\pi\nu\sigma)\right)\Delta}{\nu^2 r^2} \end{bmatrix}$$

Verify that Amat(0) = identity.

**(c56)** amat(0)

**(d56)**
$$\begin{bmatrix}1 & 0 & 0 & 0\\ 0 & 1 & 0 & 0\\ 0 & 0 & 1 & 0\\ 0 & 0 & 0 & 1\end{bmatrix}$$

Verify that Amat satisfies the defining differential equation for parallel translation.

**(c57)** (ratsubst(c_2^2 - c_1^2, nu^2 , subst(theta=%pi/2,
2*%pi*ratsubst(r^2+a^2*cos(theta)^2,rho^2,mcs_cffmat[4])) . amat(sigma)),
trigsimp(ratsimp( diff(amat(sigma), sigma) + %%, \delta)) = zeromatrix(4,4) )

**(d57)**
$$\begin{bmatrix}0 & 0 & 0 & 0\\ 0 & 0 & 0 & 0\\ 0 & 0 & 0 & 0\\ 0 & 0 & 0 & 0\end{bmatrix} = \begin{bmatrix}0 & 0 & 0 & 0\\ 0 & 0 & 0 & 0\\ 0 & 0 & 0 & 0\\ 0 & 0 & 0 & 0\end{bmatrix}$$

The acceleration, pulled back to the starting point equator(0) by parallel translation, is:

**(c58)** (ratsubst(c_2^2 - c_1^2, nu^2, amat(-sigma) . covdiffuuframe2 ),
acceleration0 : factor(apply1(%%, delta_expand)) )

**(d58)**
$$\begin{bmatrix}\dfrac{4\,\pi^2\,a\,k2\,\nu\,(r^2+m\,r-q^2)\,\sin(2\,\pi\,\nu\,\sigma)}{r^3}\\ -\dfrac{\left(4\,\pi^2\,k2\,(r^6-2\,m\,r^5+q^2\,r^4-2\,a^2\,m\,r^3+2\,a^2\,q^2\,r^2-a^2\,m^2\,r^2+2\,a^2\,m\,q^2\,r-a^2\,q^4)\ast\cos(2\,\pi\,\nu\,\sigma)\right)}{r^6}\\ 0\\ -\dfrac{4\,\pi^2\,k2\,\nu\,\sqrt{r^2-2\,m\,r+q^2+a^2}\,\sin(2\,\pi\,\nu\,\sigma)}{r}\end{bmatrix}$$

## 3.3 Development into Minkowski Space

Let

* x(t) map [0, 1] into a Riemannian manifold M.
* tanmap be isometry from tangent space at x(0) to tangent space at a point y(0) in N (Minkowski space).

Definition "Development of a Curve": Development of x(t) is a curve y(t) in N

* starting at y(0) with velocity [tanmap($A^{-1}$(t) . x'(t))]
* with acceleration [tanmap($A^{-1}$(t) . x''(t))]. (A is parallel translation in M.)

Intuittively: y(t) reflects the acceleration of x(t) in M but not the curvature of M.

Define an isometry from the tangents at equator(0) in Kerr space to the tangents at a point in Minkowski space.

**(c59)** tanmap: [c_2/nu, 0,0, c_1/nu ; 0,1,0,0 ; -c_1/nu, 0,0, -c_2/nu ; 0,0, 1, 0]

**(d59)**

$$\begin{bmatrix} -\frac{\sqrt{\Delta}}{\nu\, r} & 0 & 0 & \frac{a\left(-r+\frac{q^2}{r}-m\right)}{\nu\, r^2} \\ 0 & 1 & 0 & 0 \\ -\frac{a\left(-r+\frac{q^2}{r}-m\right)}{\nu\, r^2} & 0 & 0 & \frac{\sqrt{\Delta}}{\nu\, r} \\ 0 & 0 & 1 & 0 \end{bmatrix}$$

Tanmap depends on Kerr coordinate r and parameters a, m and q. Tanmap preserves the Minkowski metric $LFG_{ab}$.

**(c60)** ratsimp(apply1( subst(c_2^2 - c_1^2, nu^2, tanmap^^` . lfg . tanmap), delta_expand))

**(d60)**

$$\begin{bmatrix} -1 & 0 & 0 & 0 \\ 0 & 1 & 0 & 0 \\ 0 & 0 & 1 & 0 \\ 0 & 0 & 0 & 1 \end{bmatrix}$$

The acceleration along equatorial path equator(σ), mapped by tanmap to the development Minkowski space N, is:

**(c61)** (tanmap . acceleration0,
factor( apply1(%%/nu^2, delta_expand) ),
ratsubst(nu^2, apply1(c_2^2 - c_1^2, delta_expand), %%),
acceleration1 : ratsimp(subst( apply1(c_2^2 - c_1^2, delta_expand), nu^2, %%))*nu^2 )

**(d61)**

$$\begin{bmatrix} 0 \\ -4\,\pi^2\,k2\,\nu^2\cos(2\,\pi\,\nu\,\sigma) \\ -4\,\pi^2\,k2\,\nu^2\sin(2\,\pi\,\nu\,\sigma) \\ 0 \end{bmatrix}$$

Since Minkowski space is flat with affine coordinates, parallel translation is trivial.

**(c62)** (yvec: genvector(lambda([k], funmake(concat('y,k-1),[sigma])),4),
eqn_flat: diff(yvec, sigma,2) = acceleration1)

**(d62)**

$$\begin{bmatrix} y0(\sigma)_{\sigma\sigma} \\ y1(\sigma)_{\sigma\sigma} \\ y2(\sigma)_{\sigma\sigma} \\ y3(\sigma)_{\sigma\sigma} \end{bmatrix} = \begin{bmatrix} 0 \\ -4\,\pi^2\,k2\,\nu^2\cos(2\,\pi\,\nu\,\sigma) \\ -4\,\pi^2\,k2\,\nu^2\sin(2\,\pi\,\nu\,\sigma) \\ 0 \end{bmatrix}$$

The first order initial conditions for the developed curve $y(\sigma)$ are:

**(c63)** initcond_flat1 : diff(yvec, sigma) = factor(apply1(tanmap . uframe, delta_expand))

**(d63)**

$$\begin{bmatrix} y0(\sigma)_\sigma \\ y1(\sigma)_\sigma \\ y2(\sigma)_\sigma \\ y3(\sigma)_\sigma \end{bmatrix} = \begin{bmatrix} -\dfrac{2\pi a\left(3 m r^3 - 2 q^2 r^2 + a^2 m r - a^2 q^2\right)}{\nu r^4} \\ 0 \\ \dfrac{2\pi\sqrt{r^2 - 2 m r + q^2 + a^2}\left(r^4 - a^2 m r + a^2 q^2\right)}{\nu r^4} \\ 0 \end{bmatrix}$$

Restate the initial conditions in terms of coefficients k1 and k2.

**(c64)** initcond_flat2 : apply1(initcond_flat1, k1_def, k2_def, delta_def)

**(d64)**

$$\begin{bmatrix} y0(\sigma)_\sigma \\ y1(\sigma)_\sigma \\ y2(\sigma)_\sigma \\ y3(\sigma)_\sigma \end{bmatrix} = \begin{bmatrix} -k1 \\ 0 \\ 2\pi\, k2\, \nu \\ 0 \end{bmatrix}$$

Apply the first order initial conditions.

**(c65)** for k thru 4 do atvalue(part(initcond_flat2,1,k,1), sigma=0, part(initcond_flat2,2,k,1))$

The initial point in Minkowski space is arbitrary. These initial values exploit symmetry about the axis $(x, y) = (0, 0)$.

**(c66)** (map(lambda([yy,yval], atvalue(funmake(yy,[sigma]),sigma=0, yval)), [y0,y1,y2,y3], [0,k2,0,0]), printprops([y0,y1,y2,y3], atvalue) )$

$$y0(0) = 0$$

$$\left. y0(@1)_{@1} \right|_{[@1 = 0]} = -k1$$

$$y1(0) = k2$$

$$\left. y1(@1)_{@1} \right|_{[@1 = 0]} = 0$$

$$y2(0) = 0$$

$$\left. y2(@1)_{@1} \right|_{[@1 = 0]} = 2\pi\, k2\, \nu$$

$$y3(0) = 0$$

$$\left. y3(@1)_{@1} \right|_{[@1 = 0]} = 0$$

Find the solution to the equations for the developed curve.

**(c67)** odematsys(eqn_flat, yvec)

**(d67)**
$$\left[\begin{bmatrix} y0(\sigma) \\ y1(\sigma) \\ y2(\sigma) \\ y3(\sigma) \end{bmatrix} = \begin{bmatrix} -k1\ \sigma \\ k2\ \cos(2\pi\ \nu\ \sigma) \\ k2\ \sin(2\pi\ \nu\ \sigma) \\ 0 \end{bmatrix}\right]$$

Write the solution as a Macsyma function.

**(c68)** soln_flat(sigma) := ''(part(%,1,2))

**(d68)**
$$soln_flat(\sigma) := \begin{bmatrix} -k1\ \sigma \\ k2\ \cos(2\pi\ \nu\ \sigma) \\ k2\ \sin(2\pi\ \nu\ \sigma) \\ 0 \end{bmatrix}$$

Verify that the solution satisfies the differential equations for development in flat space.

**(c69)** diff(soln_flat(sigma), sigma,2) = rhs(eqn_flat)

**(d69)**
$$\begin{bmatrix} 0 \\ -4\pi^2\ k2\ \nu^2\ \cos(2\pi\ \nu\ \sigma) \\ -4\pi^2\ k2\ \nu^2\ \sin(2\pi\ \nu\ \sigma) \\ 0 \end{bmatrix} = \begin{bmatrix} 0 \\ -4\pi^2\ k2\ \nu^2\ \cos(2\pi\ \nu\ \sigma) \\ -4\pi^2\ k2\ \nu^2\ \sin(2\pi\ \nu\ \sigma) \\ 0 \end{bmatrix}$$

Verify that the solution satisfies the first order initial conditions.

**(c70)** at(diff(soln_flat(sigma), sigma), sigma=0)

**(d70)**
$$\begin{bmatrix} -k1 \\ 0 \\ 2\pi\ k2\ \nu \\ 0 \end{bmatrix}$$

Evaluate the end point of the loop.

**(c71)** at(soln_flat(sigma), sigma=0)

**(d71)**
$$\begin{bmatrix} 0 \\ k2 \\ 0 \\ 0 \end{bmatrix}$$

## 3.4 Translational Holonomy

Translational holonomy is defined as (final point - initial point) of the developed curve in Minkowski space.

**(c72)** kerr_holonomy1 : factor(soln_flat(1) - soln_flat(0))

**(d72)**
$$\begin{bmatrix} -k1 \\ k2\ (\cos(2\pi\ \nu) - 1) \\ k2\ \sin(2\pi\ \nu) \\ 0 \end{bmatrix}$$

Frequency ν <~ 1, so define α = 1 - ν >~ 0 to use in trig functions.Compute the gradiant of α with respect to r.

**(c73)** (alpha_exp : 1 - nu_exp, gradef(alpha, r, ratsubst(alpha, alpha_exp, diff(alpha_exp, r))), 'alpha = alpha_exp)

**(d73)**

$$\alpha = 1 - \sqrt{-\frac{2\,m}{r} + \frac{q^2}{r^2} - \frac{2\,a^2\,m}{r^3} + \frac{a^2\left(2\,q^2 - m^2\right)}{r^4} + \frac{2\,a^2\,m\,q^2}{r^5} - \frac{a^2\,q^4}{r^6} + 1}$$

Substitute 1- α for ν in holonomy.

**(c74)** kerr_holonomy2 : subst(nu = 1- alpha, kerr_holonomy1)

**(d74)**

$$\begin{bmatrix} -k1 \\ \left(\cos(2\,\pi\,\alpha) - 1\right) k2 \\ -\sin(2\,\pi\,\alpha)\,k2 \\ 0 \end{bmatrix}$$

Expand k1 and k2 to express holonomy in terms of m, r, a, q, and α.

**(c75)** kerr_holonomy3 :
ratsimp(subst(nu = 1- alpha, apply1( kerr_holonomy2, k1_expand, k2_expand, delta_expand)), r)

**(d75)**

$$\begin{bmatrix} \dfrac{2\,\pi\,a\left(3\,m\,r^3 - 2\,q^2\,r^2 + a^2\,m\,r - a^2\,q^2\right)}{(\alpha - 1)\,r^4} \\ \dfrac{\left(\cos(2\,\pi\,\alpha) - 1\right)\sqrt{r^2 - 2\,m\,r + q^2 + a^2}\left(r^4 - a^2\,m\,r + a^2\,q^2\right)}{(\alpha - 1)^2\,r^4} \\ -\dfrac{\sin(2\,\pi\,\alpha)\sqrt{r^2 - 2\,m\,r + q^2 + a^2}\left(r^4 - a^2\,m\,r + a^2\,q^2\right)}{(\alpha - 1)^2\,r^4} \\ 0 \end{bmatrix}$$

Assume |a| << r, m << r, q = 0. In this case, α ~ m/r ~ 0.

Set small quantities --> 0 where they are added to larger quantities, as determined by comparing powers of r.

**(c76)** kerr_holonomy4 : subst(3*r^2+a^2 = 3*r^2, factor(subst([q=0, alpha=0], kerr_holonomy3)))

**(d76)**

$$\begin{bmatrix} -\dfrac{6\,\pi\,a\,m}{r} \\ 0 \\ 0 \\ 0 \end{bmatrix}$$

### 3.5 Approximate Torsion for an Isolated Kerr Mass

Approximate torsion of an isolated Kerr mass as translational holonomy per unit area of the equatorial loop.

**(c77)** kerr_approx_torsion1 : factor(kerr_holonomy2/ (%pi*r^2))

**(d77)**

$$\begin{bmatrix} -\dfrac{k1}{\pi\, r^2} \\ \dfrac{(\cos(2\,\pi\,\alpha)-1)\,k2}{\pi\, r^2} \\ -\dfrac{\sin(2\,\pi\,\alpha)\,k2}{\pi\, r^2} \\ 0 \end{bmatrix}$$

Express approximate Kerr torsion in terms of m, r, a, q, and $\alpha$.

**(c78)** kerr_approx_torsion2 : factor(subst(nu = 1- alpha, apply1(kerr_approx_torsion1, k1_expand, k2_expand, delta_expand)))

**(d78)**

$$\begin{bmatrix} \dfrac{2\,a\,(3\,m\,r^3 - 2\,q^2\,r^2 + a^2\,m\,r - a^2\,q^2)}{(\alpha-1)\,r^6} \\ \dfrac{(\cos(2\,\pi\,\alpha)-1)\sqrt{r^2 - 2\,m\,r + q^2 + a^2}\,(r^4 - a^2\,m\,r + a^2\,q^2)}{\pi\,(\alpha-1)^2\,r^6} \\ -\dfrac{\sin(2\,\pi\,\alpha)\sqrt{r^2 - 2\,m\,r + q^2 + a^2}\,(r^4 - a^2\,m\,r + a^2\,q^2)}{\pi\,(\alpha-1)^2\,r^6} \\ 0 \end{bmatrix}$$

Assume $|a| << r$, $m << r$. In this case, $\alpha \sim m/r \sim 0$.

We wisih to compare our results with standard results from gravittational theory that include m and a but not q. In order to control expression complexity, we accomplish this by setting $q = 0$ where we want to compare with standard results in gravitational theory.

Set small quantities --> 0 where they are added to larger quantities, as determined by comparing powers of r.

**(c79)** kerr_approx_torsion3 : subst(3*r^2 + a^2 = 3*r^2, factor(subst([q = 0, alpha = 0], kerr_approx_torsion2)))

**(d79)**

$$\begin{bmatrix} -\dfrac{6\,a\,m}{r^3} \\ 0 \\ 0 \\ 0 \end{bmatrix}$$

# Part Two: Continuum Limit of a Distribution of Discrete Rotating Masses

This computer algebra script was created with Macsyma 2.4.1.a using the component tensor package.

## 4. Holonomy and Torsion in the Continuum Limit

### 4.1 Holonomy in the Continuum Limit

Consider a sequence of regular distributions of Kerr masses that approaches a continuum limit with constant densities of mass, angular momentum, and charge. Let r be the average half-distance between adjacent Kerr masses. As r becomes smaller, m and q decrease as $r^3$, and a does not change. (Recall m, a and q are properties of each Kerr mass.)

Comment: Perhaps a better value for r is $(6/\pi)^{(1/3)}$ * (half-distance between Kerr masses). This value makes the volume of the sphere of radius r equal to the volume of the cube in flat space that should be allocated to each Kerr mass. When we derive torsion as (translational holonomy)/area, this extra factor makes no significant difference because the numerator and denominator are each multiplied by (approximately) the same factor. For the same reason, any adjustment to the value of r to account for the curvature caused by the interaction of Kerr masses also makes no significant difference in this ratio.

Express the quantities computed above in terms of

$m_{density} = m/(4\pi/3\ r^3)$ = density of matter in continuum limit.

Substitute $md = m/r^3$ in computations to simplify expressions.

a = (angular momentum)/mass, which is unchanged in the continuum limit

$q_{density} = q\ (4\pi/3\ r^3)$ = density of electric charge in continuum limit.

Substitute $qd = q/r^3$ in computations to simplify expressions.

In Planck units, $1\ kg/m^3 \sim 2 * 10^{-97}$, so assume $0 < md << 1$.

**(c80)** (assume(md > 0, md < 1), density_substitutions : [m = md*r^3, q = qd*r^3])

**(d80)**
$$[m = md\ r^3, q = qd\ r^3]$$

Express $\alpha$ in terms of continuum densities and store as a transformation rule.

**(c81)** (alpha_expc: subst(density_substitutions, expand(alpha_exp)), apply(defrule, ['alpha_expandc, alpha, alpha_expc]), disprule(alpha_expandc))

**(d81)** alpha_expandc(defrule) : $\alpha \rightarrow$1
$$-\sqrt{-a^2\ qd^4\ r^6 + 2\ a^2\ md\ qd^2\ r^4 + qd^2\ r^4 + 2\ a^2\ qd^2\ r^2 - a^2\ md^2\ r^2 - 2\ md\ r^2 - 2\ a^2\ md + 1}$$

Express the approximate torsion in terms of md, a, qd, r, and $\alpha$.

**(c82)** cont_holonomy1 : ratsimp(combine(multthru(subst(nu=1- alpha, subst(density_substitutions, apply1(kerr_holonomy2, k1_expand, k2_expand, delta_expand))))), a, r, qd, md)

**(d82)**

$$\begin{bmatrix} \dfrac{2\pi a\left(qd^2\left(-2r^4 - a^2 r^2\right) + md\left(3r^2 + a^2\right)\right)}{\alpha - 1} \\ -\dfrac{\left(\cos(2\pi\alpha) - 1\right)\left(-a^2 qd^2 r^2 + a^2 md - 1\right)\sqrt{qd^2 r^6 - 2\,md\,r^4 + r^2 + a^2}}{(\alpha-1)^2} \\ \dfrac{\sin(2\pi\alpha)\left(-a^2 qd^2 r^2 + a^2 md - 1\right)\sqrt{qd^2 r^6 - 2\,md\,r^4 + r^2 + a^2}}{(\alpha-1)^2} \\ 0 \end{bmatrix}$$

Assume $md\, r^2 << 1$ and $a^2 << r^2$, in which case $\alpha \sim md\, r^2 \sim 0$.

**(c83)** cont_holonomy2 : subst(2*r^2+a^2 = r^2,ratsimp(subst([alpha = 0, 3*r^2 + a^2=3*r^2], cont_holonomy1), r, qd))

**(d83)**

$$\begin{bmatrix} 2\pi a r^2\left(qd^2 r^2 - 3\,md\right) \\ 0 \\ 0 \\ 0 \end{bmatrix}$$

We can normally assume $qd^2 r^2 << md$: in Planck units, $1\ C/m^3 \sim 2 \times 10^{-87}$, $1\ kg/m^3 \sim 2 \times 10^{-97}$, and $10^{22} \sim r$.

The holonomy is proportional to the area of the loop ($r^2$).

**(c84)** cont_holonomy3 : subst(qd^2 * r^2 - 3 * md = -3 * md, cont_holonomy2)

**(d84)**

$$\begin{bmatrix} -6\pi a\, md\, r^2 \\ 0 \\ 0 \\ 0 \end{bmatrix}$$

**Alternate case**: Assume $r^2 << a^2$ so we can take the limit r --> 0. This limit restricts the argument to classical physics.

**(c85)** cont_holonomy2x : ratsimp(subst([3*r^2 + a^2 = a^2, -2*r^4 - a^2*r^2= -a^2*r^2, alpha=0], cont_holonomy1), r)

**(d85)**

$$\begin{bmatrix} 2\pi a^3\left(qd^2 r^2 - md\right) \\ 0 \\ 0 \\ 0 \end{bmatrix}$$

**Alternate case** (continued): Assume $r^2 << a^2$ and $qd^2 r^2 << md$ as above.

Holonomy is proportional to $a^3$ md, and not to the area of the loop.

**(c86)** cont_holonomy3x : subst(qd^2 * r^2 - md = - md, cont_holonomy2x)

**(d86)**

$$\begin{bmatrix} -2\pi a^3\, md \\ 0 \\ 0 \\ 0 \end{bmatrix}$$

## 4.2 Torsion in the Continuum Limit

Compute approximate torsion of a single Kerr mass in the distrbution of Kerr masses as holonomy/($\pi$ $r^2$), where $\pi$ $r^2$ ~ area of the loop as seen from infinity.

**(c87)** cont_torsion1 : cont_holonomy3 / (%pi*r^2)

**(d87)**
$$\begin{bmatrix} -6\, a\, md \\ 0 \\ 0 \\ 0 \end{bmatrix}$$

The definition of md in terms of m omits a factor of 4/3 $\pi$, so md = $m_{density}$ 4/3 $\pi$, where $m_{density}$ = continuum mass density.

Similarly, qd = $q_{density}$ 4/3 $\pi$, where $q_{density}$ = continuum charge density.

**(c88)** cont_torsion2 : ratsimp(subst(md = m[density] *4*%pi/3, cont_torsion1))

**(d88)**
$$\begin{bmatrix} -8\, \pi\, a\, m_{density} \\ 0 \\ 0 \\ 0 \end{bmatrix}$$

## 4.3 Torsion in Coordinate Frame

Express holonomy in coordinate frame [ d/dt, d/dr, d/d$\theta$, d/d$\phi$ ] using frame field $FR^{\mu}{}_{a}$ .

**(c89)** fr_mat : subst([theta=%pi/2, qd = 0, r^2+a^2=r^2, -2*md*r^4 + r^2 + a^2=r^2, a/r=0, a/r^2=1/r], subst(density_substitutions, apply1(fr, delta_expand, rho_expand)))

**(d89)**
$$\begin{bmatrix} 1 & 0 & 0 & 0 \\ 0 & 1 & 0 & 0 \\ 0 & 0 & \frac{1}{r} & 0 \\ \frac{1}{r} & 0 & 0 & \frac{1}{r} \end{bmatrix}$$

Express holonomy vector in coordinate frame [ d/dt, d/dr, d/d$\theta$, d/d$\phi$ ] using frame field $FR^{\mu}{}_{a}$ , in limit r -> infinity.

**(c90)** coord_torsion1 : limit(fr_mat . cont_torsion2, r, inf)

**(d90)**
$$\begin{bmatrix} -8\, \pi\, a\, m_{density} \\ 0 \\ 0 \\ 0 \end{bmatrix}$$

# 5. Spin Density and the Spin-Torsion Equation

The Einstein-Cartan equation relating spin and torsion is:

**(c91)** torsion = 8*%pi * \k * "spin_density"

**(d91)**
$$\text{torsion} = 8\, \pi\, K\, \text{spin_density}$$

Torsion is a 3-index tensor whose natural indices have these properties.

* two lower spacetime (Greek) indices that specify the plane of a loop traversed to measure translation,

* a third (upper) frame (Roman) index that specifies the direction of translation.

Any index can change between spacetime and frame index using the frame field, and raised or lowered using the metric.

Below, the sign of the angular momentum is negative because (a) the spin tensor is naturally a (0,2) tensor, (b) in the test frame of a fermion, torsion has one timelike upper index; and (c) g[time, time] < 0.

Using the volume as seen by an observer at infinity, the r-$\phi$ spin density of a Kerr mass for a neighborhood of radius r is:

**(c92)** kerr_spin_density : [- a*m[density] ; 0; 0; 0]

**(d92)**
$$\begin{bmatrix} - a\, m_{density} \\ 0 \\ 0 \\ 0 \end{bmatrix}$$

Using the continuum approximations for torsion and spin density, and K = 1, we get :

**(c93)** coord_torsion1 = 8 * %pi * kerr_spin_density

**(d93)**
$$\begin{bmatrix} - 8\, \pi\, a\, m_{density} \\ 0 \\ 0 \\ 0 \end{bmatrix} = \begin{bmatrix} - 8\, \pi\, a\, m_{density} \\ 0 \\ 0 \\ 0 \end{bmatrix}$$

## Appendix E: Spacetime and fiber tensor indices

Each tensor index in continuum mechanics represents either a conserved current or a spacetime direction. Antisymmetrized spacetime indices represent flux boxes through which current flows are measured. Connection forms and curvatures also fit this pattern, though the spacetime boxes measure holonomy, not flux.

- "Spacetime indices" ($\lambda$, $\mu$, $\nu$, …), represent spacetime directions. Derivatives with respect to these indices occur only in exterior derivatives, divergences (Hodge duals of exterior derivatives), and Lie derivatives (such as strain ~ Lie derivative of a metric with respect to displacement). Spacetime indices are not covariant differentiated. In GR, it is often convenient to apply the Levi-Civita connection, which yields the same result as coordinate derivatives.
- "Fiber indices" ($a, b, c$… for vectors, $A, B, C$… for spinors, $P, Q, R$… for complex fiber spaces), represent conserved currents. No derivatives are taken with respect to these indices. These indices are covariant-differentiated with the full connection including affine torsion.

Here are some examples of the distinction between base space and fiber indices.

- Momentum $P_a{}^\mu$ has fiber index $a$ for linear momentum and index $\mu$ for a spacetime hypersurface.
- Spin density $J_{ab}{}^\mu$ has antisymmetric fiber indices $a$, $b$ for intrinsic a.m. and spacetime index $\mu$.
- The electromagnetic potential $i\,A_\mu$ and Yang-Mills potentials $i\,A^P{}_{Q\mu}$ are connection coefficients with spacetime index $\mu$ and antisymmetric indices $P$, $Q$ for a unitary rotation. $i\,A_\mu$ has an unwritten pair of indices for the unique plane in the representation space of group $U(1)$.
- The electromagnetic field $i\,F_{\mu\nu}$ and Yang–Mills fields $i\,F^P{}_{Q\mu\nu}$ are curvatures.
- A Dirac spinor $\psi^A$ has a spinor space fiber index $A$.

## Appendix F: Mathematics of connections on fiber bundles

F.1 *Mathematical notation*

$\Xi$ a smooth (pseudo-) Riemannian manifold of dimension $n$ with local coordinates $\xi^\mu$, metric $g_{\mu\nu}$, and metric connection 1-form $\omega$: $TP(\Xi, G) \to L(G)$, connection coefficients $\Gamma_{\lambda\nu}{}^\mu(\sigma)$ and covariant differentiation $D$. Denote a point by $\xi \in \Xi$.

$T(\Xi)$ the tangent bundle of the manifold $\Xi$

$GL(n)$ linear frame group of the linear frame bundle of $\Xi$

$Y$ a linear space $R^n$ with linear coordinates $y^i$ on which $GL(n)$ acts on the left

$P(\Xi, GL(n))$ the linear frame bundle over $\Xi$ with structure group $GL(n)$, with bundle projection $\pi$: $P \to \Xi$.

$B(\Xi,GL(n),Y)$ the bundle associated with $P(\Xi,GL(n))$ with fiber $Y$, with bundle projection $\pi$: $P \to \Xi$

$A(n)$ affine group that is bundle-isomorphic to the affine frame bundle of $\Xi$

$X$ an affine space $A^n$ with affine coordinates $x^i$ on which $A(n)$ acts on the left

$P(\Xi,A(n))$ a principal bundle over $\Xi$ with structure group $A(n)$. We do not use the affine frame bundle of $\Xi$, because the affine frame bundle has a fixed solder form, and we shall vary the solder form.

$B(\Xi,A(n),X)$ the bundle associated with $P(\Xi,A(n))$ with fiber $X$.

$\Xi'$ is a flat smooth (pseudo-) Riemannian manifold of dimension $n$, with local coordinates $\xi'^\mu$, metric $g'_{\mu\nu}$, metric connection 1-form $\omega$, connection coefficients $\Gamma'_{\lambda\nu}{}^\mu(\sigma)$ and covariant differentiation $D'$. Denote a point by $\xi' \in \Xi'$.

$C: [0,1] \to \Xi$ a smooth curve in $\Xi$, with $C(0) = \xi_0$ and tangent vector field $\underline{\mathbf{u}}$.

$C': [0,1] \to \Xi'$ a smooth curve in $\Xi'$, with $C'(0) = \xi'_0$ with tangent vectorfield $\underline{\mathbf{u}}'$

$L: T_\xi\Xi \to T_{\xi'}\Xi'$ is a linear isometry between tangent spaces at $\xi \in \Xi$ and $\xi' \in \Xi'$.

F.2. *Connections, curvature and torsion on fiber bundles*

We assume the reader is familiar with the theory of connections on fiber bundles [Bishop and Crittenden 1964; Kobayashi and Nomizu 1963 and 1969]. This section focuses on the relation between holonomy and curvature.

The most general and elegant definition of holonomy is given in terms of the theory of connections on fiber bundles. Let $P$ be a principal bundle over $\Xi$ with structure group $G$, endowed with a connection. Define a smooth curve $C$ in $\Xi$ starting and ending at the point $\xi$. Starting at a point $p_0$ in the fiber over $\xi$, form the horizontal lift of the curve $C$ into $P$. The lifted curve will end at a point $p_1$ in the fiber over $\xi$. Then there is a unique element $g \in G$ such that

(F-1) $$p_0\, g = p_1$$

By the equivariance property of connections, $g$ is independent of the choice of $p_0$, and depends only on the connection on $P$ and the curve $C$ in $\Xi$.

$g$ is the holonomy of the loop $C$. If we have a linear representation of the structure group, then the holonomy is a transformation from any basis of the representation space by parallel translation around $C$.

The most general and elegant definition of curvature is given in terms of the theory of connections on fiber bundles. There are two equivalent ways to construct the curvature:

(i) Choose a point $\xi \in \Xi$ and two vectors $v$ and $w$ at $\xi$. Construct a one-parameter family of closed loops $C_t(\sigma)$ through $\xi$ which are tangent at $\xi$ to the plane of $v \wedge w$, and which converge to $\xi$ as $t \to 0$. Let $g_t$ be the holonomy of the loop $C_t$. Then the curvature $R$ is defined as

(F-2) $$-R(v \wedge w) = \lim_{t\to 0} \frac{g_t}{\text{area}(t)}$$

where $\text{area}(t)$ = area of the loop $C_t$. The curvature depends only on the 2-form $v \wedge w$, and not on the choice of the vectors $v$ and $w$ or the curves $C_t$. The limit exists and is finite when appropriate smoothness conditions are imposed upon the connection. This construction yields a simple interpretation of curvature as holonomy per unit area, in the limit $\text{area}(t) \to 0$.

(ii) Let $\omega$ = the connection 1-form on the bundle $P$.

(F-3) $$\omega: TP \to L(G)$$

The curvature $\Omega$ is the horizontal component of the 2-form $d\omega$ (exterior derivative of $\omega$):

(F-4) $$\Omega(\cdot\,,\ \cdot) = d\omega\,(\text{Hor}(\cdot), \text{Hor}(\cdot))$$

where $\text{Hor}$ is the horizontal projection map for tangents to the bundle $P$.

The horizontal 2-form $\Omega$ on $P$ uniquely determines a Lie-algebra-valued 2-tensor field on $\Xi$. If the bundle is a linear frame bundle, then the Lie algebra components of the curvature can be identified with $(1, 1)$ rotation tensor fields on $\Xi$. This yields a $(1, 3)$ tensor field $R$ on $\Xi$, which is the curvature tensor usually used in Riemannian geometry.

The second definition of curvature is the customary one. The first one offers richer intuitive insights for the purposes of this paper. We can derive the first definition of curvature from the second using the relation

(F-5) $$d\omega(X, Y) = (X\,\omega(Y) - Y\,\omega(X) - \omega([X, Y]))$$

which holds for all 1-forms ω and all smooth vector fields X and Y. If ω is the connection form and X and Y are the horizontal lifts of vector fields on Ξ, then

(F-6) $$d\omega(X, Y) = -\omega([X, Y])$$

For the X and Y specified, [X, Y] is vertical, and ω is a vertical projection operator (identifying the Lie algebra of the structure group with the vertical fiber). [X, Y] is defined by traversing an integral curve of X followed by those of Y, –X, and –Y, subtracting the coordinates of the initial location from those of the end point given by the integration process, and taking the limit, per unit area enclosed, as the lengths of integral curves traversed approach zero. When projected into Ξ, this construction for[X, Y] yields a family of closed loops at the point $p \in \Xi$, like the loops used in the first definition of curvature. The construction for [X, Y] amounts to parallel translation around these closed loops, and ω([X, Y]) is the holonomy per unit area in the limit of small areas of the loops traversed. The conventional normalization for R is that

(F-7) $$R(v, w) = 2\, d\omega(X, Y)$$

where v and w are projections in Ξ of the vectors X and Y tangent to P. Hence R is minus the holonomy per unit area.

F.3. *Relation between development, curvature, and torsion for linear connections*

Linear connections are characterized by two conditions:

(i) The structure group G is a subgroup of the linear automorphisms of $R^n$ onto itself (commonly represented by $GL(n, R)$ = group of n-by-n real matrices).

(ii) There is a solder 1–form : $TP \rightarrow R^n$ which is horizontal and equivariant. This form identifies a point $p \in P$ with a basis of tangents to Ξ at $\pi(p)$.

There are two ways to define the torsion of a linear connection:

(i) At a point $\xi \in \Xi$, form a family of loops $C_t(\sigma)$ which converge to p and are tangent to $v \wedge w$, as in Section 3.2 above. Develop the curves $C_t$ into a flat (pseudo-) Euclidean n manifold. The developed curves generally do not close. Torsion can be defined as

(F-8) $$-T(v \wedge w) = \lim_{t \rightarrow 0} \frac{(C'_t(1) - C'_t(0))}{\text{area}(t)}$$

Torsion is translational holonomy (the failure-to-close vector) per unit area enclosed by the loop, in the limit as the area approaches zero. It is translational curvature. In the theory of affine connections, where the structure group is a subgroup of the inhomogeneous automorphisms of $R^n$, torsion appears as the curvature components associated with translations in the Lie algebra of the structure group.

(ii) Torsion can be defined as the horizontal part of the exterior derivative of the fundamental 1-form θ:

(F-9) $$\text{Tor}(\cdot, \cdot) = d\theta(\text{Hor}(\cdot), \text{Hor}(\cdot))$$

The horizontal 2-form Tor on B determines a unique (1, 2) tensor field T on Ξ, by transforming the $R^n$–valued index of Tor into tangent vectors of Ξ. T is the torsion tensor in Riemann-Cartan geometry.

The second definition of torsion can be used to derive the first definition in terms of translational holonomy, much the same as in the case of curvature. This construction requires use of the inhomogeneous linear group as the structure group.

**Appendix G: Generation of rotational and translational holonomy on flat manifolds**

We will derive rotational holonomy and translational holonomy from a flat space. We employ the methods of "surgery on manifolds: cutting and gluing edges of manifolds, and including and excising pieces of manifolds. Surgery has been used in mathematics since the 1960s to alter algebraic topological invariants of manifolds [Milnor 1963]. In this section we use surgery to generate rotational holonomy which, in the continuum limit of many small defects, yields rotational curvature, and translational holonomy which, in the continuum limit of many small defects, yields translational curvature (affine torsion).

A crystal lattice is a discrete version of an affine manifold. The concepts used below have been used in metallurgy and crystallography since the 1950s [Kondo 1955; Bilbo 1957].

a) Derive a disclination (metallurgists' terminology; "in-plane rotational holonomy" to geometers) from a flat 2-D affine manifold. Figure (G..**1**–a) is a graphic representation of a disclination before it is glued closed.
   i) Start with a flat disk. Cut the disk from center to edge.
   ii) Excise an angular sector with central angle $\theta$.
   iii) Connect the two boundary cuts to form a cone.

   The tip of the cone has a Dirac delta of positive Riemannian curvature, as can be seen in two ways:

   - Parallel translation around the tip of the cone yields rotational holonomy of angle $\theta$.
   - A disk of radius $r$ around the tip of the cone has an area deficit of $\frac{1}{2}\, \theta\, r^2$.

   If you insert (instead of excise) a sector with central angle $\theta$, the center of the disk has a Dirac delta of negative Riemannian curvature.

   If you distribute $n$ small excisions of angle $\theta/n$ evenly in the area of the disk, you have a discrete approximation to space of positive Riemannian curvature.

b) Derive a screw dislocation (metallurgists' terminology; "orthogonal translational holonomy" to geometers) from a flat 3-D affine manifold. Figure (G..**1**–b) is a graphic representation of screw dislocation.

   i) Start with a 3-D cube with Cartesian coordinates $-1 \leq x, y, z \leq 1$.
   ii) Cut the cube from center to the edge along the surface $0 \leq x \leq 1$ and $y = z = 0$.
   iii) Re-attach the cut surfaces, except translate one surface upward (or downward) by distance $\Delta z$.

   The center of the cube has a screw dislocation with Burghers vector (metallurgists' terminology; translational holonomy to geometers) that is orthogonal to the plane of the loop path, with magnitude $\Delta z$ in the $z$-direction. This can be seen by parallel translating a vector around a path in the plane $z = 0$; that path does not close because there is no plane $z = 0$.

   In EC, a screw dislocation describes the angular momentum of a small classical rotating object.

c) Derive an edge dislocation (metallurgists' terminology; 'in-plane translational holonomy" to geometers) from a flat 2-D or 3-D manifold. Figure (G..**1**–c) is a 3–D graphic representation of an edge dislocation.

   i) Start with square or cube that has Cartesian coordinates $x = 0$, $y, z \in [-2, 2]$.
   ii) Cut the square or cube from the center to the edge along $x \in [-2, 2]$, $-2 \leq y \leq 0$.
   iii) Excise a rectangular strip of width $\Delta z$ from one face of the cut.
   iv) Reconnect corresponding points on the two cut faces.

   The center of the square has an edge dislocation in the plane of the paper ($y$, $z$ plane) with Burghers vector (metallurgists' terminology; translational holonomy to geometers) that is orthogonal to the cut.

   This can be seen as follows:

   - Parallel translate a vector around a square path in the $y$, $z$ having sides $10\, \Delta z$.
   - The path will not close by the amount $\Delta z$ in the $z$–direction. This is the translational holonomy.

**Figure G.1: Graphic representations of disclination, edge dislocation, and screw dislocation**

| **(a) Disclination**<br>Discrete unit of<br>positive Riemannian curvature | **(b) Screw dislocation**<br>Discrete unit of<br>classical angular momentum | **(c) Edge dislocation**<br>Discrete unit of torsion trace |
|---|---|---|
| 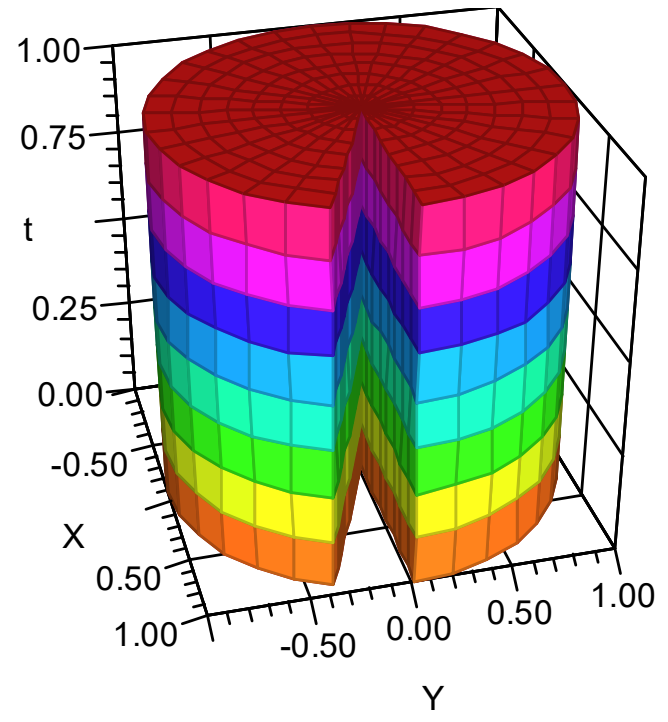 | 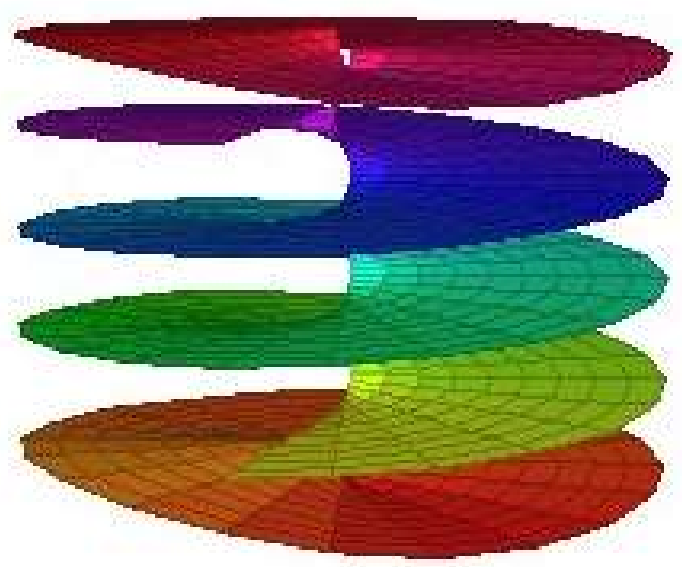 | 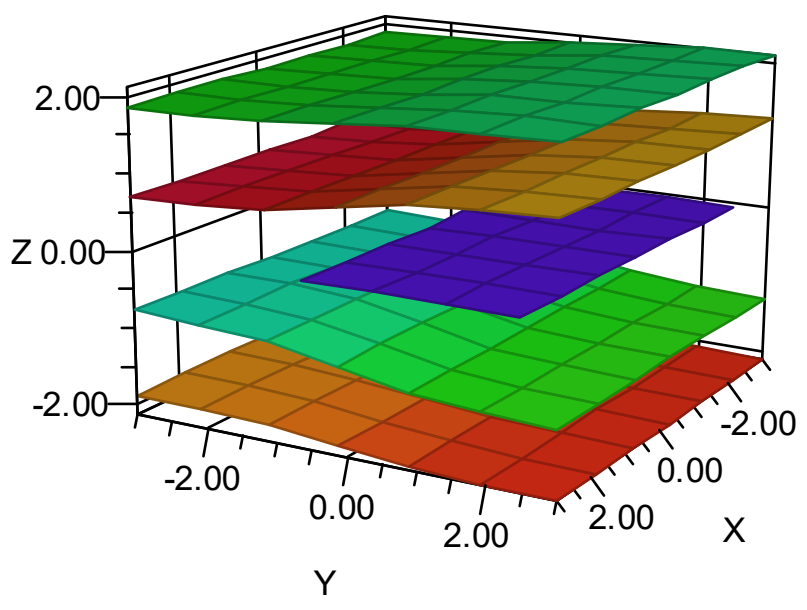 |

At the top and bottom of Figure (G..**1**-b), the screw dislocation ends and creates two edge dislocations – edges of a sheet that are not connected to anything, like the free edge in Figure (G..**1**-c). This is an illustration of conservation of dislocation lines in crystals, which becomes the conservation of intrinsic angular momentum in EC [Petti 2001].

Surgery of manifolds and discrete affine defects are used in this section to make the geometry clear to those who are not familiar with it. The Riemannian geometry of the GR solutions contains translational holonomy, and the limit of this holonomy for very small loops is affine torsion.

## Acknowledgements

I would like to thank R K Sachs for nurturing my knowledge of relativity and an inquisitive attitude many years ago. Friedrich Hehl generously contributed his time to improve this work in several ways: he corrected inaccurate historical comments; he pointed out where arguments and explanations should be strengthened; and he contributed to a vigorous discussion of the limitations of rigor of the continuum computation. Jean Krisch provided encouragement and helpful suggestions. Jean de Valpine brought to my attention the correspondence between Einstein and Cartan.